\documentclass[%
 reprint,
superscriptaddress,
 amsmath,amssymb,
 aps,
]{revtex4-2}

\usepackage{graphicx}
\usepackage{epstopdf}
\usepackage{epsfig}
\usepackage{subcaption}
\usepackage{dcolumn}
\usepackage{bm}
\usepackage{caption}
\makeatletter
\renewcommand\@makecaption[2]{%
  \par
  \vskip\abovecaptionskip
  \begingroup
   \small\rmfamily
    \begingroup
     \samepage
     \flushing
     \let\footnote\@footnotemark@gobble
     \@make@capt@title{#1}{#2}\par
    \endgroup
  \endgroup
  \vskip\belowcaptionskip
}
\makeatother

\begin{document}

\preprint{APS/123-QED}

\title{Charging Dynamics of Electrical Double Layers Inside a Cylindrical Pore: Predicting the Effects of Arbitrary Pore Size}

\author{Filipe Henrique}
\affiliation{Department of Chemical and Biological Engineering, University of Colorado, Boulder}
\author{Pawel J. Zuk}%
\affiliation{%
Institute of Physical Chemistry, Polish Academy of Sciences,  Warsaw, Poland}%
\affiliation{%
Department of Physics, Lancaster University, Lancaster, United Kingdom}%
\author{Ankur Gupta}
\email{Corresponding author:
ankur.gupta@colorado.edu}
\affiliation{Department of Chemical and Biological Engineering, University of Colorado, Boulder}

\date{\today}

\begin{abstract}
Porous electrodes are found in energy storage devices such as supercapacitors and pseudocapacitors. However, the effect of electrode-pore-size distribution over their energy storage properties remains unclear. Here, we develop a model for the charging of electrical double layers inside a cylindrical pore for arbitrary pore size. We assume small applied potentials and perform a regular perturbation analysis to predict the evolution of electrical potential and ion concentrations in both the radial and axial directions. We validate our perturbation model with direct numerical simulations of the  Poisson-Nernst-Planck equations, and obtain quantitative agreement between the two approaches for small and moderate potentials. Our analysis yields two main characteristic features of arbitrary pore size: i) a monotonic decrease of the charging timescale with an increase in relative pore size (pore size relative to Debye length); {ii) large potential changes for overlapping double layers in a thin transition region, which we approximate mathematically by a jump discontinuity.} {We quantify the contributions of electromigration and charge diffusion fluxes which provide mechanistic insights into the dependence of charging timescale and capacitance on pore size.} We develop a modified transmission circuit model that captures the effect of arbitrary pore size and demonstrate that a time-dependent {transition-region resistor} needs to be included in the circuit. We also derive {phenomenological} expressions for {average} effective capacitance and charging timescale as a function of pore-size distribution. {We} show that the capacitance and charging timescale increase {with smaller average pore sizes and with smaller polydispersity}, resulting in a gain of energy density at a constant power density. Overall, our results advance the {mechanistic} understanding of electrical-double-layer charging.
\end{abstract}

\maketitle


\section{Introduction}

Batteries and fuel cells are traditional porous-material-based electrochemical devices. {Over the past 60 years,} new devices such as supercapacitors \cite{becker1957} have started to emerge. They are comprised of porous electrodes -- typically made up of dispersions of activated carbon spheres -- immersed in aqueous, organic, or ionic liquid electrolytes. The electrodes store charge through the physical adsorption of dissociated ions onto their pore surfaces, forming a charged region commonly referred to as the electrical double layer \cite{biesheuvel2010nonlinear,zhang2014highly}. Obviating the need for redox reactions, these devices charge faster than batteries and present better cyclability. Nevertheless, their energy density and capacitance are limited by their available specific surface area \cite{zhang2014highly}. In view of these characteristics, supercapacitors bridge the gap between traditional capacitors and batteries, being used in situations where fast response and moderate energy output are required, e.g., stabilization of energy fluctuations in power grids \cite{presser2012electrochemical} and memory protection in electronic devices \cite{winter2004batteries}. More recently,  hybrid capacitors have been designed in an effort to utilize the energy storage mechanisms of both electrical double layers  and reduction-oxidation (redox) reactions \cite{simon2008materials,muzaffar2019review}. They consist of two distinct electrodes, one supercapacitor-like which stores charge physically into double layers, and another metal oxide pseudocapacitor-like which accumulates energy by performing fast oxidation surface reactions \cite{simon2008materials}. While there have been significant advances in the material design of supercapacitors and hybrid capacitor electrodes, the effect of pore-size distribution on the energy storage properties of these devices remains unclear \cite{simon2008materials}.

The earliest models for electrode charging address the geometry of an electrolyte between flat plates, dating back to the works of Gouy \cite{gouy1910constitution}, with later contributions from Chapman \cite{chapman1913li} and Stern \cite{stern1924theorie} to consolidate the well known Gouy-Chapman-Stern model. Notably, a wealth of different effects have been studied to complete this simplified geometrical picture of flat plate charging. Bazant et al. \cite{bazant2004diffuse} reviewed the previous flat plate charging studies and developed an approach to solve the Poisson-Nernst-Planck (PNP) equations, from the linear regime at low voltages to the nonlinear effects at high voltages. Feicht et al. \cite{feicht2016discharging} compared one-dimensional PNP solutions with the experimentally observed reverse peaks in electrolytic cell discharging. Kilic and Bazant \cite{kilic2007steric1,kilic2007steric2} derived modified PNP equations including ion-radius effects to study the influence of ion crowding in charge storage. For higher electrolyte concentrations, other effects such as multi-component diffusion given by Stefan-Maxwell fluxes \cite{balu2018role}, and ion correlations \cite{jimenez2006electrolyte,jimenez2008regimes,wu2011classical} have been studied, for instance in some continuum models \cite{gupta2018electrical,gupta2020ionic,gupta2020thermodynamics}.

The porous geometry is incorporated in the form of equivalent circuit representations, widely used in the modeling of pore charging in supercapacitor electrodes \cite{black2010pore,kaus2010modelling,kowal2011detailed,madabattula2018insights,lian2020blessing}, stemming from the pioneering work of de Levie \cite{de1963porous,de1964porous}. Later, Biesheuvel and Bazant \cite{biesheuvel2010nonlinear} extended the circuit to high potentials for capacitive deionization applications. Recent papers \cite{lian2020blessing,janssen2021transmission} further discuss the relationship between the equivalent circuit and the corresponding transmission line continuous equation for pores with finite lengths. Throughout most pore charging models, common assumptions are either of thin double layers \cite{de1963porous,de1964porous,biesheuvel2010nonlinear,biesheuvel2011diffuse} within the pores, i.e., such that the length of the charged region is much smaller than the pore size, or overlapping double layers\cite{peters2016analysis,levy2020breakdown}, where the charged regions extending from the opposite sides of the surface meet. However, pore sizes can range from less than 2 nm for micropores to more than 50 nm for macropores \cite{simon2008materials,madabattula2018insights}. Thus, a first-principles approach that extends pore charging models to arbitrary pore sizes is required in order to accurately describe the charging of supercapacitors and predict their properties, such as capacitance, energy density, and power density. Some of the works which address pore-size dependence focus on the equilibrium response \cite{varghese2011simulating}, while others propose transient transport-equation-based numerical schemes that describe porous network charging for arbitrary pore sizes \cite{alizadeh2017multiscale,alizadeh2019impact}. { However, to the best of our knowledge, the relative importance of charge transport mechanisms for arbitrary pore sizes and their implications on transmission line circuits have not been reported.}

\par{}In our previous work \cite{gupta2020charging}, an analytical model based on the PNP equations was proposed to describe the charging of pores at low applied potentials in the limit of overlapping double layers. Here, inspired by perturbation models on electrokinetics \cite{wiersema1966calculation,o1978electrophoretic,yariv2011streaming,schnitzer2012macroscale,khair2020breaking,amrei2020perturbation}, we develop a perturbation expansion model for arbitrary pore sizes -- i.e., pore radii -- in the limit of small applied potentials. We compare the predictions of the model to direct numerical simulations to show that the perturbation solution yields good results even for moderate applied potentials ($\approx 50$ mV). We demonstrate that a modified transmission line circuit (compared to classic supercapacitor literature \cite{de1963porous,de1964porous}) includes a {resistor representing finite changes in} electric potential { at a thin entrance region} at the mouth of a pore. We derive an effective capacitance to study the effect of pore-size distribution on energy and power densities. Besides addressing transient charging, {the solution developed here quantifies the contributions of diffusion and electromigration and provides mechanistic insights into the charging process of pores of arbitrary sizes}.

\section{Problem Formulation}
\begin{figure}[h!]
    \centering
    \includegraphics[scale=1.0]{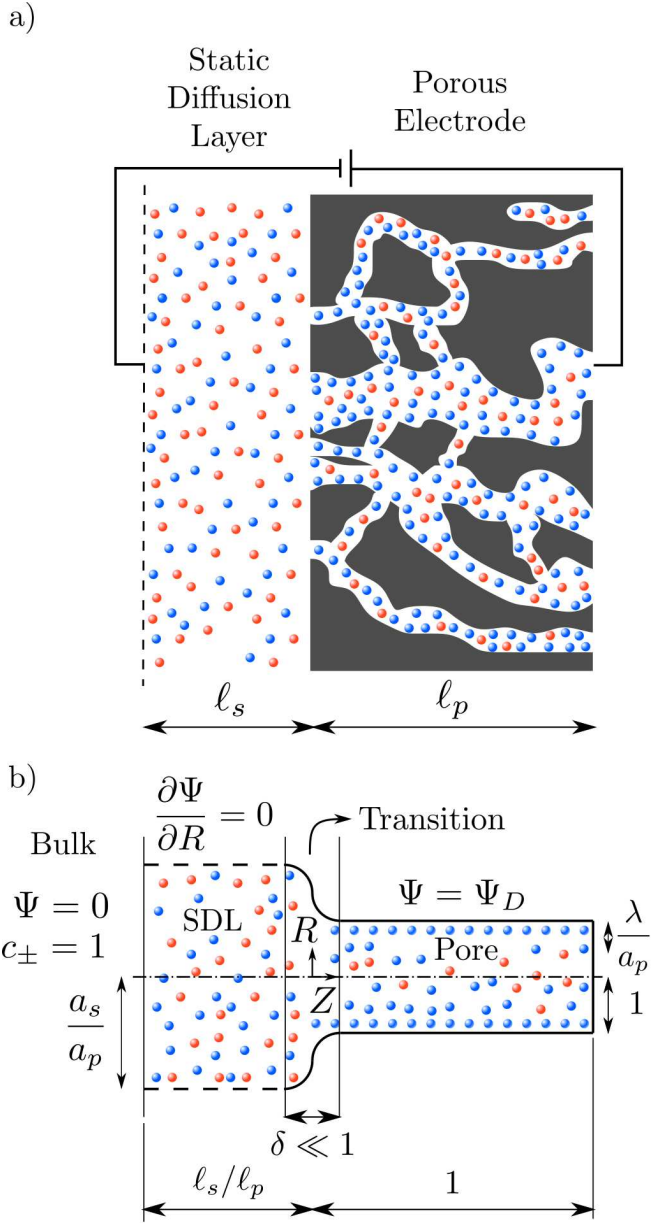}
    \caption{{A schematic of the problem setup. a) A porous electrode is subjected to an applied voltage. The electrode pores are filled with an electrolyte with cations (in red) and anions (in blue). The pore electrolyte is in contact with a static diffusion layer. b) Schematic representation of a single cylindrical pore (using dimensionless variables) and the division of the domain into bulk, static diffusion layer (SDL), transition, and pore regions. The bulk has $\Psi=0$ and equal concentration of both ionic species $c_\pm=1$. The SDL radial boundary, with vanishing normal gradients, is represented by dashed lines. The potential at the surface of the pore is $\Psi_D$, the lengths of the static diffusion layer and pore are $\ell_s/\ell_p$ and $1$, respectively. The radii of SDL and the pore are $a_s/a_p$ and $1$, respectively, and the Debye length is $\lambda/a_p$.} }
    \label{Fig: schematic}
\end{figure}

We consider { an ideally conducting} porous electrode that consists of tortuous pores of different radii, filled with an electrolyte; see the schematic in Fig. \ref{Fig: schematic}a. Once a voltage difference is applied, counterions are attracted to the pore surfaces, forming electrical double layers of thicknesses that may be thin or comparable to the pore size \cite{zhang2014highly,boota2015graphene}. We follow common practice \cite{bazant2004diffuse,biesheuvel2010nonlinear, gupta2020charging} in assuming the formation of a static diffusion layer (SDL), an electroneutral region beside the electrode. Fig. \ref{Fig: schematic}b shows a simplified setup with a cylindrical pore of length $\ell_p$ and radius $a_p$, where the radial and axial directions are denoted by $r$ and $z$, respectively. The SDL, of length $\ell_s$ and radius $a_s$, is adjacent to the pore. We denote the cation and anion concentrations by $c_{\pm}(r,z,t)$ and electric potential by $\psi(r,z,t)$, where $t$ is time. The position $z=0$ represents the interface between the pore and the SDL. We assume that the SDL is in contact with the bulk such that  $c_{\pm}(r,-\ell_s,t)=c_0$ and $\psi (r,-\ell_s,t)=0$. At $t=0$, the concentration of ions everywhere is $c_{\pm}(r,z,0)=c_0$. At $t>0$, {since the electrode material is an ideal conductor,} the potential at the surface of the pore $\psi(a_p, z, t)=\psi_D$. We also assume that the pore surface is ideally blocking, i.e., the flux of ions across the pore surface is zero.  In addition, we assume that $\frac{a_p}{\ell_p} \ll 1$, $\frac{\ell_s}{\ell_p} = O(1)$, the ions are monovalent, that ion diffusivities inside the pore are equal and given by $D_p$. The ion diffusivities outside the pore are also assumed equal and are given by $D_s$. We note that the ion diffusivities inside and outside the pore may be different due to confinement effects. As we show later, the ratio of diffusivities -- i.e., $\frac{D_p}{D_s}$ -- dictates the interaction between the pore and the SDL. We denote Debye length by $\lambda $ such that $\lambda=\sqrt{\frac{\varepsilon k_B T}{2 e^2 c_0}}$, where $\varepsilon$ is the electrical permittivity. For reference, a 1 M aqueous electrolyte at room temperature has a Debye length $\lambda\approx 0.3$ nm. The objective of this article is to determine the value of $\psi(r,z,t)$ and $c_{\pm}(r,z,t)$ for an arbitrary relative pore size $\frac{a_p}{\lambda}$.\par{}

\par{} Physically, when the potential is applied on the surface of a pore, oppositely charged ions migrate inside the pore, while the similarly charged ions transport out of the pore. This relative transport of ions produces an electrical current. In addition, due to the charge imbalance, an electrical double layer forms adjacent to the surface. As time progresses, the potential drop across the electrical double layer starts to saturate, ions stop migrating, and the current ceases. 
\par{} To solve for $\psi$ and $c_{\pm}$, we start by writing the Poisson-Nernst-Planck equations \cite{deen1998analysis,bard2000electrochemical,newman2012electrochemical}
\begin{subequations}
\begin{eqnarray}
\frac{\partial c_{\pm}}{\partial t} + \nabla \cdot \mathbf{N}_{\pm} = 0, \\
-\varepsilon \nabla^2 \psi  = e(c_+ - c_-), \label{Eq: poisson}
\end{eqnarray}
where $\mathbf{N}_{\pm}$ are the ion fluxes. Inside the pore
\begin{equation}
\mathbf{N}_{\pm} = -D_p \nabla c_{\pm} \mp \frac{D_p e c_{\pm}}{k_B T} \nabla \psi, \label{Eq: Np}
\end{equation}
and outside the pore
\begin{equation}
\mathbf{N}_{\pm} = -D_s \nabla c_{\pm} \mp \frac{D_s e c_{\pm}}{k_B T} \nabla \psi.
\end{equation}
\label{Eq: pnp}
\end{subequations}
We note that $\mathbf{N}_{\pm}= \mathbf{e}_r N_{\pm r}  + \mathbf{e}_z N_{\pm z} $ by angular symmetry.  We introduce dimensionless charge density $\rho = \frac{c_+ - c_-}{c_0}$, salt concentration $s = \frac{c_+ + c_-}{c_0}$, potential $\Psi = \frac{e \psi}{k_B T}$, time $\tau = \frac{t D_p}{\ell_p^2}$, axial position $Z=\frac{z}{\ell_p}$, radial position $R=\frac{r}{a_p}$, and gradient operator $\bar{\nabla}=\ell_p\nabla$. Note that $-\frac{\ell_s}{\ell_p} \le Z \le 1 $. In addition, $0 \le R \le 1 $ for $ 0 \le Z \le 1$ (i.e., the pore region), and $0 \le R \le \frac{a_s}{a_p} $ for $ -\frac{\ell_s}{\ell_p} \le Z < 0$ (i.e., the static diffusion layer region; see Fig. \ref{Fig: schematic}b). With these substitutions, the set of Eqs. (\ref{Eq: pnp}) becomes
\begin{subequations}
\begin{align}
\frac{\partial \rho}{\partial \tau} + \bar{\nabla} \cdot \mathbf{J} = 0,  \label{Eq: dim_charge} \\
\frac{\partial s}{\partial \tau} + \bar{\nabla} \cdot \mathbf{W} = 0, \label{Eq: dim_salt} \\
- \left( \frac{\lambda}{\ell_p} \right)^2 \bar{\nabla}^2 \Psi  = \frac{\rho}{2}, \label{Eq: dim_poisson}
\end{align}
where $\bar{\nabla} = \mathbf{e}_R \left( \frac{\ell_p}{a_p} \right) \frac{\partial}{\partial R}  +  \mathbf{e}_Z \frac{\partial}{\partial Z}$, $\mathbf{J} = \frac{\mathbf{N}_+ - \mathbf{N}_-}{D_p c_0/\ell_p}$ is the dimensionless charge flux and $\mathbf{W} = \frac{\mathbf{N}_+ + \mathbf{N}_-}{D_p c_0/\ell_p}$ is the dimensionless salt flux. Inside the pore
\begin{equation}
\mathbf{J} = - \left( \bar{\nabla} \rho +  s \bar{\nabla} \Psi \right), \mathbf{W} = - \left( \bar{\nabla} s + \rho \bar{\nabla} \Psi \right), \label{Eq: dim_Np}
\end{equation}
and outside the pore
\begin{equation}
\mathbf{J} = - \frac{D_s}{D_p} \left( \bar{\nabla} \rho +  s \bar{\nabla} \Psi \right), \mathbf{W} =  - \frac{D_s}{D_p} \left( \bar{\nabla} s + \rho \bar{\nabla} \Psi \right). \label{Eq: dim_Ns}
\end{equation}
\label{Eq: dim_pnp}
\end{subequations}
At $\tau=0^+$, inside the pore, the potential is constant and equal to the wall potential since the electrical double layer hasn't developed yet. In contrast, the potential is linear in the SDL due to electroneutrality. Therefore, Eqs. (\ref{Eq: dim_pnp}) are subjected to the following initial conditions
\begin{subequations}
\begin{align}
&\rho(R,Z,0^+)=0, \\
& s(R,Z,0^+)=2, \\ 
&\Psi(R,Z<0,0^+)=\Psi_D \left( 1 + \frac{ Z \ell_p }{\ell_s} \right),  \\
&\Psi(R,Z \ge 0,0^+)=\Psi_D.
 \end{align}
 \label{eq: ics}
\end{subequations}
\noindent  \par{} The boundary conditions (BCs) at $Z=-\frac{\ell_s}{\ell_p}$ are given by the bulk condition, or
\begin{subequations}
\begin{align}
&\rho(R,- \frac{\ell_s}{\ell_p},\tau)=0,\\
&s(R,- \frac{\ell_s}{\ell_p},\tau)=2,\\
&\Psi(R,- \frac{\ell_s}{\ell_p},\tau)=0.
\label{Eq: bcsbulk3}
\end{align}
\label{Eq: bcsbulk}
\end{subequations}
\par{} At the end of the pore, i.e., $Z=1$, the BCs are that gradients of potential and concentration vanish for $l_p\gg a_p$, implying
{
\begin{equation}
   \left. \frac{\partial \rho}{\partial Z} \right|_{Z=1}=\left. \frac{\partial s}{\partial Z} \right|_{Z=1}=\left. \frac{\partial \Psi}{\partial Z} \right|_{Z=1} = 0.
\label{eq: bcsend}
\end{equation}}
Due to the symmetry, the BCs at the center of the system, i.e., $R=0$, are simply 
{ \begin{equation}
\left. \frac{\partial \rho}{\partial R} \right|_{R=0}=\left. \frac{\partial s}{\partial R} \right|_{R=0}=\left. \frac{\partial \Psi}{\partial R} \right|_{R=0}=0.
\label{Eq: bcsymm}
\end{equation}}

At the surface of the pore, i.e., $R=1$ and $Z \ge 0$, the BCs are of ideally blocking electrode and constant potential, which are given by
\begin{subequations}
\begin{eqnarray}
\left. J_R \right|_{R=1}=0, \\
\left. W_R \right|_{R=1}=0, \\
\Psi(1,Z>0, \tau)=\Psi_D. 
 \end{eqnarray}
 \label{Eq: bcporesurf}
 \end{subequations}
{ We note that Eq. (\ref{Eq: bcporesurf}c) represents the ideally conducting electrode BC}. Finally, at the boundaries of the SDL, i.e.,  $R=\frac{a_s}{a_p}$ and $Z<0$, the BCs are vanishing gradients, or
{
\begin{equation}
\left. \frac{\partial \rho}{\partial R} \right|_{R=\frac{a_s}{a_p}}=\left. \frac{\partial s}{\partial R} \right|_{R=\frac{a_s}{a_p}}=\left. \frac{\partial \Psi}{\partial R} \right|_{R=\frac{a_s}{a_p}}=0.
\label{Eq: bcSDLsurf}
\end{equation}}
{This assumption is valid for non-interacting pores where the radial currents are identically zero, consistent with the treatment of SDL in the literature \cite{biesheuvel2010nonlinear,gupta2020charging,janssen2021transmission}.} We solve the set of Eqs. (\ref{Eq: dim_pnp}) -- (\ref{Eq: bcSDLsurf}) numerically using OpenFOAM \cite{weller1998tensorial,jasak2007openfoam}. The details of geometry, mesh, and algorithm have been described in Ref. \cite{gupta2020charging}. We refer to the solution from OpenFOAM as direct numerical simulations (DNS). Next, we perform a regular perturbation analysis \cite{deen1998analysis} to obtain an analytical expression for $\rho(R,Z,\tau)$ and $\Psi(R,Z,\tau)$.

\section{Regular Perturbation Analysis}
In this section, we focus on the small potential limit, i.e., $ | \Psi_D | \ll 1$, and conduct a regular perturbation analysis to describe the charge and potential inside the pore. To this end, we divide the solution into three regions: (I) SDL, (II) inside the pore, and (III) { transition region} between the SDL and the mouth of the pore. 
\subsection{Static Diffusion Layer}
The SDL is characterized by electroneutrality, i.e., $\rho=0$. Furthermore, for $|\Psi_D| \ll 1$, $s=2$ \cite{biesheuvel2010nonlinear, gupta2020charging}. Therefore, as per Eq. (\ref{Eq: dim_poisson}) and the boundary conditions in Eqs. (\ref{Eq: bcsbulk3}), (\ref{Eq: bcsymm}), and (\ref{Eq: bcSDLsurf}), it is straightforward to show that $\Psi$ is linear in $Z$ and is independent of $R$. Mathematically, we write 
\begin{equation}
\Psi = \Psi_{\mathrm{left}} \left( 1 + \frac{Z \ell_p}{\ell_s} \right) \ \ \textrm{for } Z < 0,
\label{Eq: Psilin}
\end{equation}
where $\Psi_{\mathrm{left}}=\Psi(R,0^{-},\tau)$, i.e., the centerline potential to the left of the SDL-pore { interface}, and is to be determined.  
\subsection{Inside the Pore}
The region inside the pore consists of ion concentration and potential varying in time as well as in both radial and axial directions. In the low-potential limit, we propose regular perturbation expansions of the dependent variables in the small parameter $\Psi_D$, i.e., 
\begin{subequations}
{
\begin{eqnarray}
\rho=\rho_0+\rho_1\Psi_D+ O(\Psi_D^2), \\
s=s_0+s_1\Psi_D+ O(\Psi_D^2), \\ 
\Psi=\Psi_0+\Psi_1\Psi_D+ O(\Psi_D^2). 
\end{eqnarray}}
We also introduce variables $\rho_{m}(Z,\tau)$ and $\Psi_{m}(Z,\tau)$ that represent the centerline charge and density. As per our proposed perturbation expansion,
{
\begin{eqnarray}
\rho_m=\rho_{m0} +\rho_{m1} \Psi_D+O(\Psi_D^2) \\
\Psi_m=\Psi_{m0}+\Psi_{m1}\Psi_D+ O(\Psi_D^2). 
\end{eqnarray} 
\label{Eq: regexp}}
\end{subequations}
\noindent In this article, we only focus on the leading-order and first-order terms.
\par{} The leading-order terms, i.e., $\rho_0$, $s_0$ and $\Psi_0$ are obtained from the response in the absence of an applied potential, i.e., $\Psi_D=0$. Thus, it can be seen that the leading-order coefficients $\rho_0=0$, $s_0=2$ and $\Psi_0=0$ satisfy the set of Eqs. (\ref{Eq: dim_pnp}) with the initial conditions (\ref{eq: ics}) and boundary conditions (\ref{eq: bcsend})--(\ref{Eq: bcporesurf}). 
\par{} Inserting the expansions provided in Eq. (\ref{Eq: regexp}) (with $\rho_0=0$, $s=2$ and $\Psi_0=0$) in Eq. (\ref{Eq: dim_charge}) and (\ref{Eq: dim_Np}) and collecting the first-order terms in $\Psi_D$ yields
\begin{equation}
    \dfrac{\partial\rho_1}{\partial \tau}=\bar{\nabla}^2\rho_1+2 \bar{\nabla}^2\Psi_1.
    \label{eq: rholowpsi}
\end{equation}
Similarly, Poisson's equation -- Eq. (\ref{Eq: poisson}) -- after substitution of the perturbation expansions takes the form
\begin{equation}
- \left( \frac{\lambda}{\ell_p} \right)^2 \bar{\nabla}^2 \Psi_1  = \frac{\rho_1}{2}.
\label{Eq: dim_poisson1}
\end{equation}
Only the known leading-order term in the salt concentration enters Eq. (\ref{eq: rholowpsi}), such that Eqs. (\ref{eq: rholowpsi}) and (\ref{Eq: dim_poisson1}) suffice to determine first-order corrections to charge density and potential in the low potential regime. Therefore, we do not need to further solve Eq. (\ref{Eq: dim_salt}).

\par{} Since  $\frac{a_p}{\ell_p} \ll 1$, the diffusion in the radial direction is much faster than in the axial direction and we can assume quasiequilibrium in the radial direction\cite{alizadeh2017multiscale}. Thus, with symmetry and ideally blocking conditions (see Eqs. (\ref{Eq: bcsymm}) -- (\ref{Eq: bcporesurf})), we have $J_{R}=0$. Utilizing this condition in Eq. (\ref{Eq: dim_Np}) and collecting the first-order terms in the perturbation expansion, we get
\begin{equation}
J_{R1} = - \frac{\partial \rho_1}{\partial R} - 2 \frac{\partial \Psi_1}{\partial R}=0,
\label{Eq: req}
\end{equation}
which implies that \cite{gupta2020charging}
\begin{equation}
\rho_1 + 2 \Psi_1 = \rho_{m1} + 2 \Psi_{m1},  
\label{Eq: req3}
\end{equation}
where $\rho_{m1}(Z,\tau)$ and $\Psi_{m1}(Z,\tau)$ are to be determined. Next, by utilizing $\frac{a_p}{\ell_p} \ll 1$ in Eq. (\ref{Eq: dim_poisson1}), we get \cite{gupta2020charging}
\begin{equation}
- \frac{1}{R} \frac{\partial}{\partial R} \left( R\frac{\partial \Psi_1}{\partial R} \right) = \left( \frac{a_p}{\lambda} \right)^2  \frac{\rho_1}{2}. \label{Eq: poissonR}
\end{equation}
Integrating Eq. (\ref{Eq: poissonR}) with $\left. \frac{\partial \Psi_1}{\partial R} \right|_{R=0} = 0$ and $\Psi_1(Z,1,\tau)=1$ yields
\begin{equation}
\frac{\Psi_1 - \Psi_{m1} - \frac{\rho_{m1}}{2}}{1 - \Psi_{m1} - \frac{\rho_{m1}}{2}} = \frac{I_0 \left(R \frac{a_p}{\lambda} \right)}{I_0 \left( \frac{a_p}{\lambda} \right)}, \label{Eq: psibessel}
\end{equation}
where $I_n$ is the modified Bessel function of the first kind of order $n$. By substituting $R=0$ in Eq. (\ref{Eq: psibessel}), we obtain
\begin{equation}
\rho_{m1} =  \frac{2 \left(\Psi_{m1} - 1 \right)}{I_0 \left( \frac{a_p}{\lambda} \right) - 1}. \label{Eq: rhombessel}
\end{equation}
Next, by combining Eqs. (\ref{Eq: psibessel}) - (\ref{Eq: rhombessel}), substituting the leading- and  first-order coefficients in Eq. (\ref{Eq: regexp}), and neglecting higher-order terms, we get
\begin{equation}
\Psi =  \Psi_{m} \left(\frac{I_0 \left( \frac{a_p}{\lambda} \right) - I_0 \left( R \frac{a_p}{\lambda} \right)}{I_0 \left( \frac{a_p}{\lambda} \right) - 1}\right)+\Psi_D\left(\dfrac{I_0 \left( R \frac{a_p}{\lambda} \right) -1}{I_0 \left( \frac{a_p}{\lambda} \right) - 1}\right). 
\label{Eq: psival}
\end{equation}
Similarly, substituting Eqs. (\ref{Eq: psibessel}) and (\ref{Eq: rhombessel}) into Eq. (\ref{Eq: req3}) and neglecting higher-order terms in the perturbation expansion of $\rho$, we get
\begin{equation}
\rho = \frac{2 \left(\Psi_{m} - \Psi_D \right) I_0 \left( R \frac{a_p}{\lambda} \right)}{I_0 \left( \frac{a_p}{\lambda} \right) - 1}. 
\label{Eq: rhoval}
\end{equation}
Next, we note that since $J_R=0$, so we can simplify Eq. (\ref{eq: rholowpsi}) to obtain
\begin{equation}
\frac{\partial \rho}{\partial \tau} = \frac{\partial^2 \rho}{\partial Z^2} + 2 \frac{\partial^2 \Psi}{\partial Z^2}. 
\label{Eq: dim_rho_ztau}
\end{equation}
To determine the axial dependence, we average Eq. (\ref{Eq: dim_rho_ztau}) over the cross-sectional area of the pore (integrating across the radial direction) by utilizing the known radial dependence in Eqs. (\ref{Eq: psival}) and (\ref{Eq: rhoval}). Mathematically,
\begin{equation}
\int_0^1 \frac{\partial \rho}{\partial \tau} R dR = \int_0^1 \frac{\partial^2 \rho}{\partial Z^2} R dR + 2 \int_0^1 \frac{\partial^2 \Psi}{\partial Z^2} R dR.
\label{Eq: int_dim_rho_ztau}
\end{equation}
\noindent We emphasize that this crucial step enables us to derive a solution for arbitrary $\frac{a_p}{\lambda}$, thereby bridging the previously reported trends in thin and overlapping double layer limits  \cite{de1963porous, de1964porous, biesheuvel2010nonlinear, gupta2020charging}. Now, substituting Eqs. (\ref{Eq: psival}) and (\ref{Eq: rhoval}) into Eq. (\ref{Eq: int_dim_rho_ztau}), we obtain
\begin{equation}
\frac{\frac{2 \lambda}{a_p} I_1 \left( \frac{a_p}{\lambda} \right) }{ I_0 \left( \frac{a_p}{\lambda} \right)} \frac{\partial \Psi_{m}}{\partial \tau} = \frac{\partial^2 \Psi_{m}}{\partial Z^2}.
\label{Eq: fin_equation}
\end{equation}
The boundary conditions for Eq. (\ref{Eq: fin_equation}) include $\Psi_{\mathrm{right}} = \Psi_{m}(0^{+},\tau)$ and $\left. \frac{\partial \Psi_{m}}{\partial Z} \right|_{Z=1}=0$, where $\Psi_{\mathrm{right}}$, i.e., the centerline potential to the right of the SDL-pore interface, is to be determined.
\subsection{ Transition Region}
{ We now focus on the transition region between the SDL and the mouth of the pore. First, we emphasize that the transition region is only relevant for overlapping double layer limits. In the thin double layer limit, $\rho=0$ inside the majority of the pore as well as the SDL region, and thus the transition region becomes irrelevant. This is consistent with the derivations in Biesheuvel et al. \cite{biesheuvel2010nonlinear, biesheuvel2011diffuse}, who assume a continuous variation in $\Psi$ across the SDL-pore interface in the thin double layer limit, thereby implicitly neglecting the transition region.
\par{} In contrast, in the overlapping double layer limit, $\rho$ inside the pore is non-zero whereas $\rho$ in the SDL region is zero, and a transition region is required to connect the two regions. Accordingly, inside the transition region, $\rho$ varies from zero to the value inside the pore. To estimate the thickness of the transition region $\delta$ (scaled by $\ell_p$), we emphasize that charge density is related to the length scale of the charge gradients by the Poisson equation; see Eq. (2c). Because the radial variation inside the SDL can be ignored, the dimensionless length scale over which charge gradients could be present inside the SDL is $\frac{\ell_s}{\ell_p}=O(1)$. Therefore, using Eq. (2c) and assuming $\frac{\lambda}{a_p} =O(1)$, it is straightforward to show that $\frac{\rho}{\Psi_D}=O(\frac{a_p}{\ell_p})^2 \ll 1$, enabling us to assume electroneutrality inside the SDL. In contrast, the smallest length scale over which the charge gradients are present inside the pore is $\frac{a_p}{\ell_p}$, which implies $\frac{\rho}{\Psi_D}=O(1)$  \cite{gupta2020charging,chen2020influence,deen1998analysis,lyklema2005fundamentals,evans1999colloidal}; see also Eq. (\ref{eq:rhosteady}). Next, in the transition region, $\rho$ varies from zero to the value inside the pore. Therefore, $\rho$ inside the transition region is on the same order of magnitude as that inside the pore. Accordingly, the relevant length scale is the same as that of the pore, or $\delta = O\left( \frac{a_p}{\ell_p} \right) \ll 1$.  In summary, the transition region is thin due to geometrical features of the pore.} 
\par{} {In the limit of $\frac{a_p}{\lambda}=O(1)$, the transition region may present finite electric potential and charge density changes \cite{gupta2020charging}. Therefore, in our perturbation expansion model, we approximate this region as a jump discontinuity. Defining $J_\mathrm{left}=\left. J_Z \right|_{Z=0^-}$ and $J_\mathrm{right}=\left. J_Z \right|_{Z=0^+}$, the charge flux in the SDL is given by $J_\mathrm{left}= - 2 \frac{D_s}{D_p} \frac{\Psi_{\mathrm{left}} \ell_p}{\ell_s}$ where $\Psi_{\mathrm{left}}=\Psi_{m}(0^{-},\tau)$; see Eqs. (\ref{Eq: dim_Ns}) and (\ref{Eq: Psilin}). In contrast, the charge flux inside the pore is given by $J_\mathrm{right}= - \left. \frac{2 I_0 \left( \frac{a_p}{\lambda} \right)}{I_0 \left( \frac{a_p}{\lambda} \right)-1}  \frac{\partial \Psi_{m}}{\partial Z} \right|_{Z=0^+}$; see Eqs. (\ref{Eq: dim_Np}), (\ref{Eq: psival}) and (\ref{Eq: rhoval}). Therefore, since the transition region is thin and thus has negligible charge storage, we ensure current conservation across it by requiring { $J_\mathrm{left}A_s=J_\mathrm{right}A_p$, i.e.,}
\begin{equation}
\frac{D_s}{D_p} \frac{ \ell_p }{\ell_s} \frac{a_s^2}{a_p^2} \Psi_{\mathrm{left}} = \left. \frac{ I_0 \left( \frac{a_p}{\lambda} \right)}{I_0 \left( \frac{a_p}{\lambda} \right)-1}  \frac{\partial \Psi_{m}}{\partial Z} \right|_{Z=0^+}.
\label{Eq: chargebal}
\end{equation}
Eq. (\ref{Eq: chargebal}) consists of two variables, i.e., $\Psi_{\mathrm{left}}$ and $\Psi_{\mathrm{right}}$, and we need an additional equation to solve $\Psi_{m}$. 
\par{} To relate $\Psi_{\mathrm{left}}$ and $\Psi_{\mathrm{right}}$, we require current conservation also on the left surface of the transition region (see Fig. 1b), relating its charge flux to that of the SDL. To do so, we define $\rho_{\mathrm{left}}=\rho_m(0^-,\tau)=0$ and $\rho_{\mathrm{right}}=\rho_m(0^+, \tau)$. { The charge flux inside the transition region can be approximated by $J_{Z, \textrm{int}}= -(\frac{\rho_{\mathrm{right}}}{\delta} + 2 \frac{\Psi_{\mathrm{right}} - \Psi_{\mathrm{left}}}{\delta})$ with a $O(\delta)$ error, where $\delta$ is the dimensionless thickness of the region, scaled by $\ell_p$. The current conservation relation{, $J_{Z,\mathrm{int}} = J_\mathrm{left}$,} reads 
\begin{equation}
    \frac{\rho_{\mathrm{right}}}{\delta} + 2 \frac{\Psi_{\mathrm{right}} - \Psi_{\mathrm{left}}}{\delta}=2 \frac{D_s}{D_p} \frac{\Psi_{\mathrm{left}} \ell_p}{\ell_s}.
\end{equation}
For $\delta\ll 1$, the right-hand side can be neglected, so} $\rho_{\mathrm{right}} + 2 \Psi_{\mathrm{right}} = 2 \Psi_{\mathrm{left}}$. Eq. (17) can then be used to derive}
\begin{equation}
    \Psi_{\mathrm{left}} =  \frac{I_0 \left( \frac{a_p}{\lambda} \right) \Psi_{\mathrm{right}}   - \Psi_D }{I_0 \left( \frac{a_p}{\lambda} \right)-1}. 
    \label{Eq: Psi1Psi2}
\end{equation}
Eq. (\ref{Eq: Psi1Psi2}) shows that when $\frac{a_p}{\lambda} \gg 1$, $\Psi_{\mathrm{left}}=\Psi_{\mathrm{right}}$, which is expected for thin double layers. In contrast, when $\frac{a_p}{\lambda} \ll 1$, $\Psi_{\mathrm{right}}=\Psi_D$, which is also expected, since the potential will be close to the surface potential everywhere inside the pore. Substituting  Eq. (\ref{Eq: Psi1Psi2}) in Eq. (\ref{Eq: chargebal}) yields 
\begin{equation} 
   \left. \frac{\partial \Psi_{m}}{\partial Z} \right|_{Z=0^+} = \textrm{Bi} \left( \Psi_{\mathrm{right}}  - \frac{\Psi_D}{I_0 \left(\frac{a_p}{\lambda} \right)} \right),
   \label{Eq: BC2}
\end{equation}
where $\mathrm{Bi}=\frac{D_s}{D_p} \frac{ \ell_p }{\ell_s} \frac{a_s^2}{a_p^2}$ {is the Biot number}. 
\subsection{Governing Equation}
We can now combine Eqs. (\ref{Eq: fin_equation}) and (\ref{Eq: BC2}) to rewrite the governing equation for centerline potential as
\begin{subequations}
\begin{equation}
    \frac{\partial \phi}{\partial T} = \frac{\partial^2 \phi}{\partial Z^2}, 
    \label{Eq: final}
\end{equation}
where $T= \frac{ I_0 \left( \frac{a_p}{\lambda} \right)}{\frac{2 \lambda}{a_p} I_1 \left( \frac{a_p}{\lambda} \right) } \tau$ and $\phi = \Psi_{m} - \frac{\Psi_D}{I_0 \left(\frac{a_p}{\lambda} \right)}$. Here, $T$ and $\phi$ are the effective dimensionless time and potential, respectively. Eq. (\ref{Eq: final}) is subjected to $\phi(Z,0)=\frac{I_0 \left( \frac{a_p}{\lambda} \right)-1}{I_0 \left( \frac{a_p}{\lambda} \right)} \Psi_D$, $ \left. \frac{\partial \phi}{\partial Z} \right|_{Z=0}=\textrm{Bi}\,\phi(0,T)$ and $\left. \frac{\partial \phi}{\partial Z} \right|_{Z=1}=0$. The analytical solution of Eq. (\ref{Eq: final}) yields \cite{deen1998analysis}
\begin{equation}
\resizebox{\linewidth}{!}{$\phi = \left( \frac{I_0 \left( \frac{a_p}{\lambda} \right)-1}{I_0 \left( \frac{a_p}{\lambda} \right)} \right) \Psi_D \sum_{n=1}^{\infty} \frac{4 \sin \kappa_n }{2 \kappa_n + \sin 2 \kappa_n} \exp(-\kappa_n^2 T) \cos (\kappa_n (Z-1))$}, 
\label{Eq: final_sol}
\end{equation}
where $\kappa_n \tan \kappa_n = \textrm{Bi}$. Eq. (\ref{Eq: final_sol}) can be used to derive $\Psi_{m} = \phi + \frac{\Psi_D}{I_0 \left(\frac{a_p}{\lambda} \right)}$, which can then be used to evaluate $\Psi$ and $\rho$ inside the pore using Eqs. (\ref{Eq: psival}) and (\ref{Eq: rhoval}). Finally, Eqs. (\ref{Eq: Psilin})  and (\ref{Eq: Psi1Psi2}) enable us to evaluate $\Psi$ inside the SDL. 
\par{} We emphasize that Eqs. (\ref{Eq: psival}), (\ref{Eq: rhoval}) and (\ref{Eq: final_sol_full}) are the key result of this paper. To the best of our knowledge, this is the first solution of the charge density and potential profiles within a cylindrical pore for arbitrary pore size in the limit of low potentials, formally justified by a regular perturbation expansion. This paper highlights that the mathematical structure of $\Psi$ remains identical irrespective of $\frac{a_p}{\lambda}$. However, the effect of $\frac{a_p}{\lambda}$ modifies the charging timescale
\begin{equation}
    t_c = \dfrac{t}{T} = \dfrac{\frac{2 \lambda}{a_p} I_1 \left( \frac{a_p}{\lambda} \right) }{ I_0 \left( \frac{a_p}{\lambda} \right)} \frac{\ell_p^2}{D_p}.
    \label{eq: tc}
\end{equation} 
\label{Eq: final_sol_full}
\end{subequations}
In the thin-double-layer limit, i.e., $\frac{\lambda}{a_p} \ll 1$, $t_c \approx \frac{2 \lambda}{a_p} \frac{\ell_p^2}{D_p}$, consistent with the results reported previously \cite{biesheuvel2010nonlinear}. In the overlapping-double-layer limit, i.e., $\frac{\lambda}{a_p} \gg 1$, $t_c \approx \frac{\ell_p^2}{D_p}$, consistent with the results reported in Gupta et. al. \cite{gupta2020charging}. {We illustrate this timescale dependence on pore size with contour plots of the time evolution of the electric potential in Fig. 2, and rationalize its mechanism in the following discussion}.\par{}
{In the thin-double-layer limit, the pore remains uncharged except in the vicinity of its surface. As shown in Fig. 2 and the Supplementary Video, for $a_p/\lambda=10$, $\tau=0.2$ displays a charge density profile very close to that of $\tau = 1$, indicating an earlier saturation of the profile and therefore a lower timescale. This is also backed up by the calculations, which predict $\tau_c \approx 0.19$ for $\frac{a_p}{\lambda}=10$; see Eq. (\ref{eq: tc}). In contrast, when double layers are not thin, i.e., $\frac{a_p}{\lambda}=2$, a radial distribution charge forms throughout the pore, producing a transient axial gradient of charge. The charge profiles continue to develop in both radial and axial directions and appear to saturate around $\tau=1$. This is also consistent with our prediction since $\tau_c \approx 0.7$; see Eq. (\ref{eq: tc}). The charging of $\frac{a_p}{\lambda}=5$ lies between the two other scenarios described. Finally, we also note that while the overlapping double layers take longer to charge, they also store more charge throughout the pore, resulting in a trade-off of charge density and charging timescale.}\par{}
{The characteristic feature that sets the charging timescale of arbitrary pore sizes is the interplay of electromigrative and diffusive fluxes. Therefore, we construct a vector plot of diffusive and electromigrative fluxes by employing our analytical derivation; see Fig. 3. First, we note that for both $\frac{a_p}{\lambda}=2$ and $\frac{a_p}{\lambda}=10$, the radial diffusive flux is always balanced by the radial electromigrative flux, even though the radial diffusive and electromigrative fluxes increase with time; see Fig. 3. This is a classical feature of double-layer charging; see Ref. \cite{bazant2004diffuse} for more details. Overall, the balance in the radial direction implies that the timescale of pore charging is controlled by the fluxes in the axial direction. For $\frac{a_p}{\lambda}=10$, the axial gradient of charge vanishes and electromigration is the only mechanism promoting axial transport of charge, in consonance with the literature \cite{de1963porous, de1964porous, biesheuvel2010nonlinear, biesheuvel2011diffuse}. In contrast, for $\frac{a_p}{\lambda}=2$, both electromigration and diffusion cooperate in driving charge transport; see Fig. 3. However, this increase in charge flux of  in the overlapping double layer limit is smaller than the boost in charge density, leading to a longer charging timescale.} 
\begin{figure*}[t]
    \centering
    \includegraphics[scale=0.36]{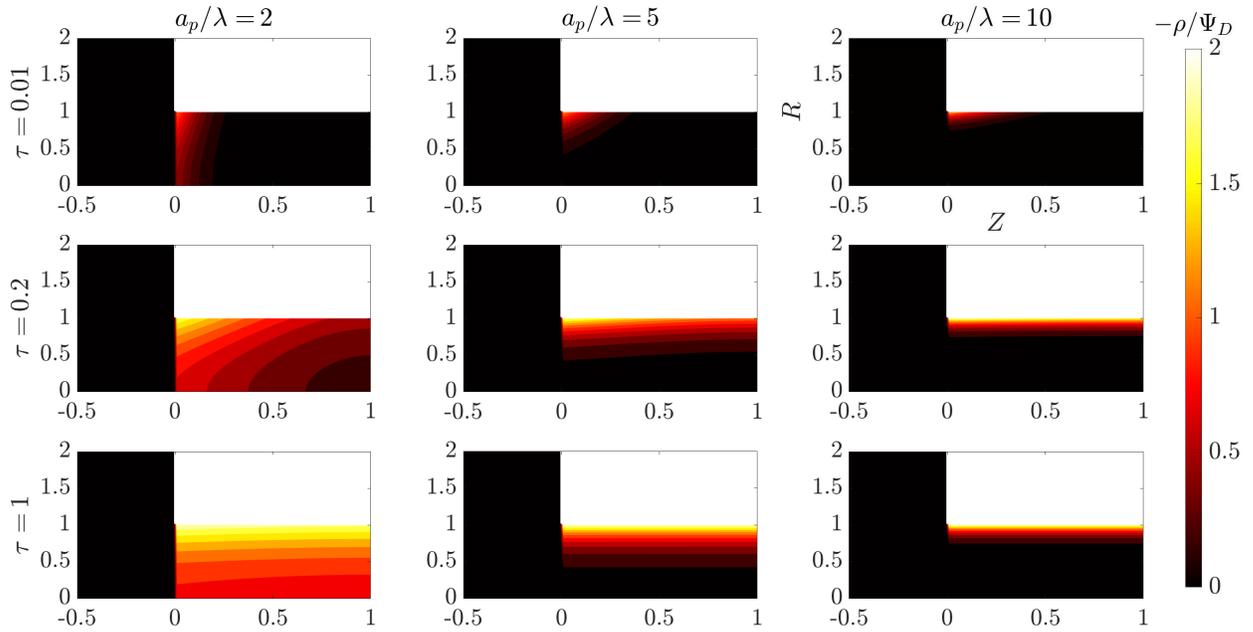}
    \caption{ Contour {plots} of the time evolution of the charge density in the SDL-pore region for different relative pore sizes. Charge density profiles reveal different charging timescales and screening lengths corresponding to different relative pore sizes. {All the plots share the same axes.}}
    \label{fig:contour}
\end{figure*}
\begin{figure*}[t]
    \centering
    \includegraphics[scale=0.32]{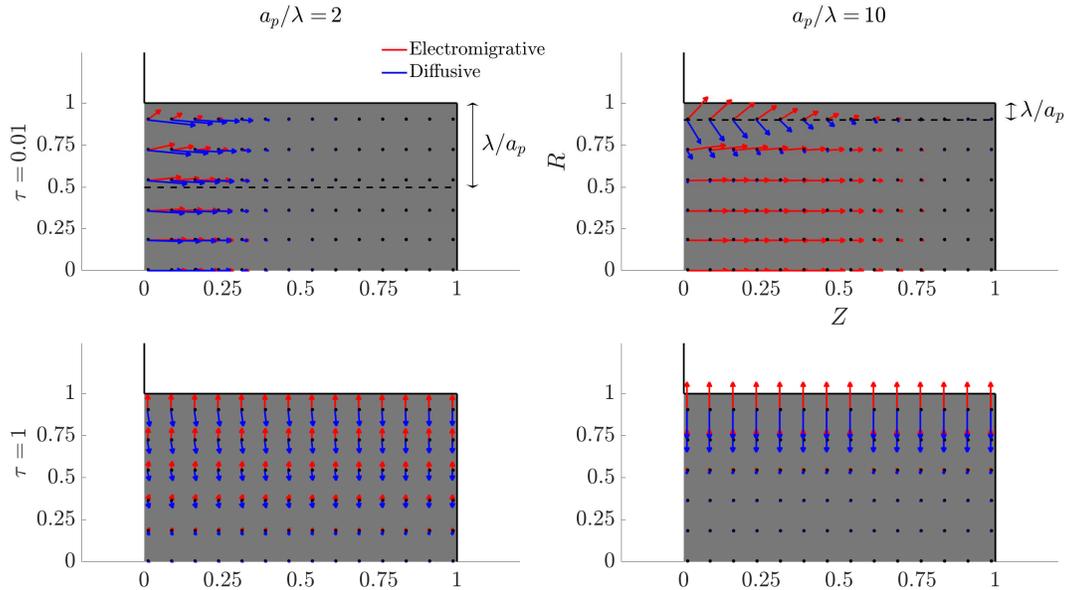}
    \caption{{Vector field {plots} of negative charge diffusive and electromigrative fluxes in the pore for different relative pore sizes. Arrow lengths are logarithmically scaled. Charge flux is only driven by electromigration for thin double layers, but set by a balance of diffusion and electromigration for overlapping double layers. {All the plots share the same axes.}}}
    \label{fig:fluxvec}
\end{figure*}

\section{Results: Potential, Charge and Current} 

\begin{figure}[h!]
    \centering
    \begin{subfigure}{0.25\textwidth}
        \caption*{\rm{a)}}
        \includegraphics[scale=0.32]{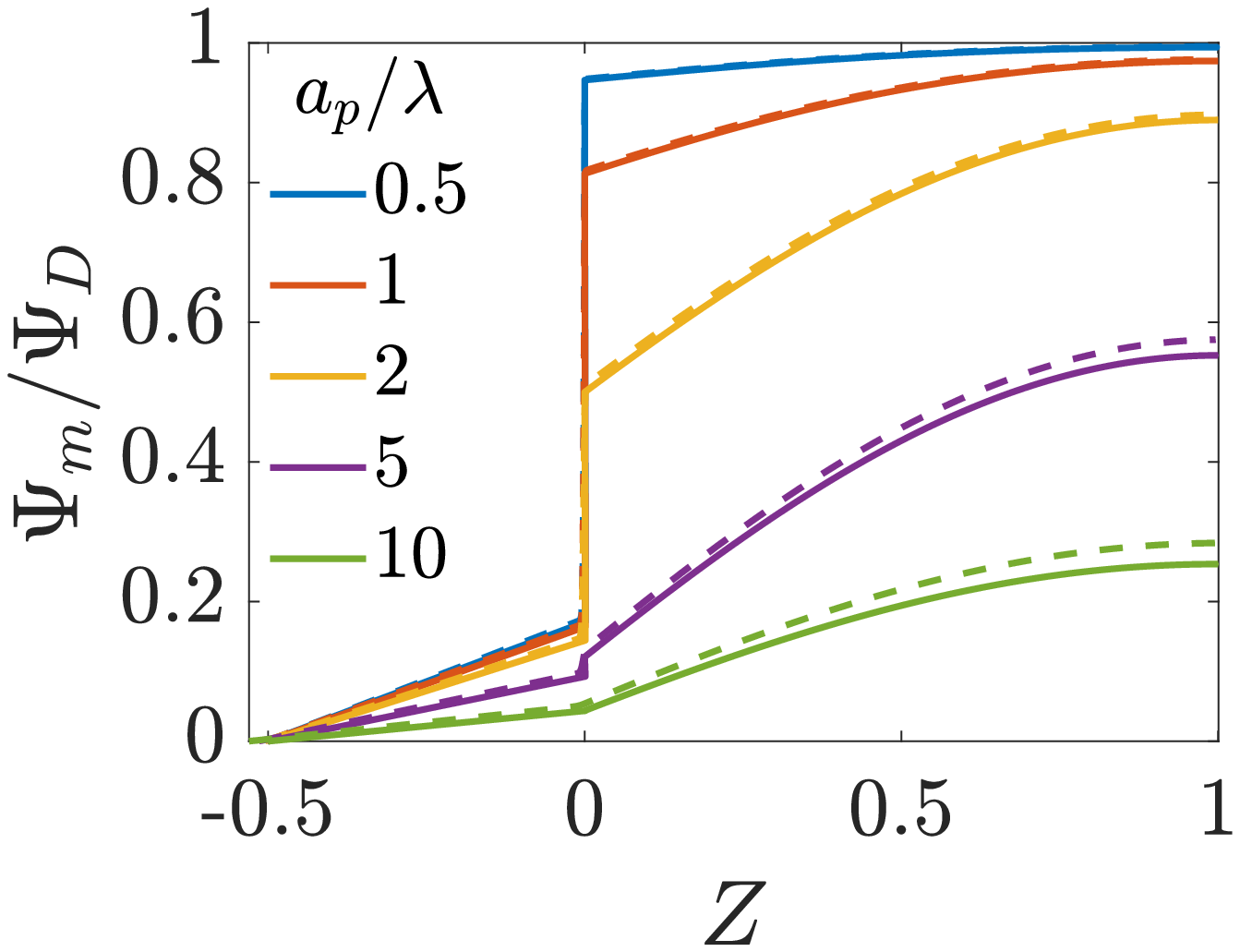}
    \end{subfigure}%
    \begin{subfigure}{0.25\textwidth}
        \caption*{\rm{b)}}
        \includegraphics[scale=0.32]{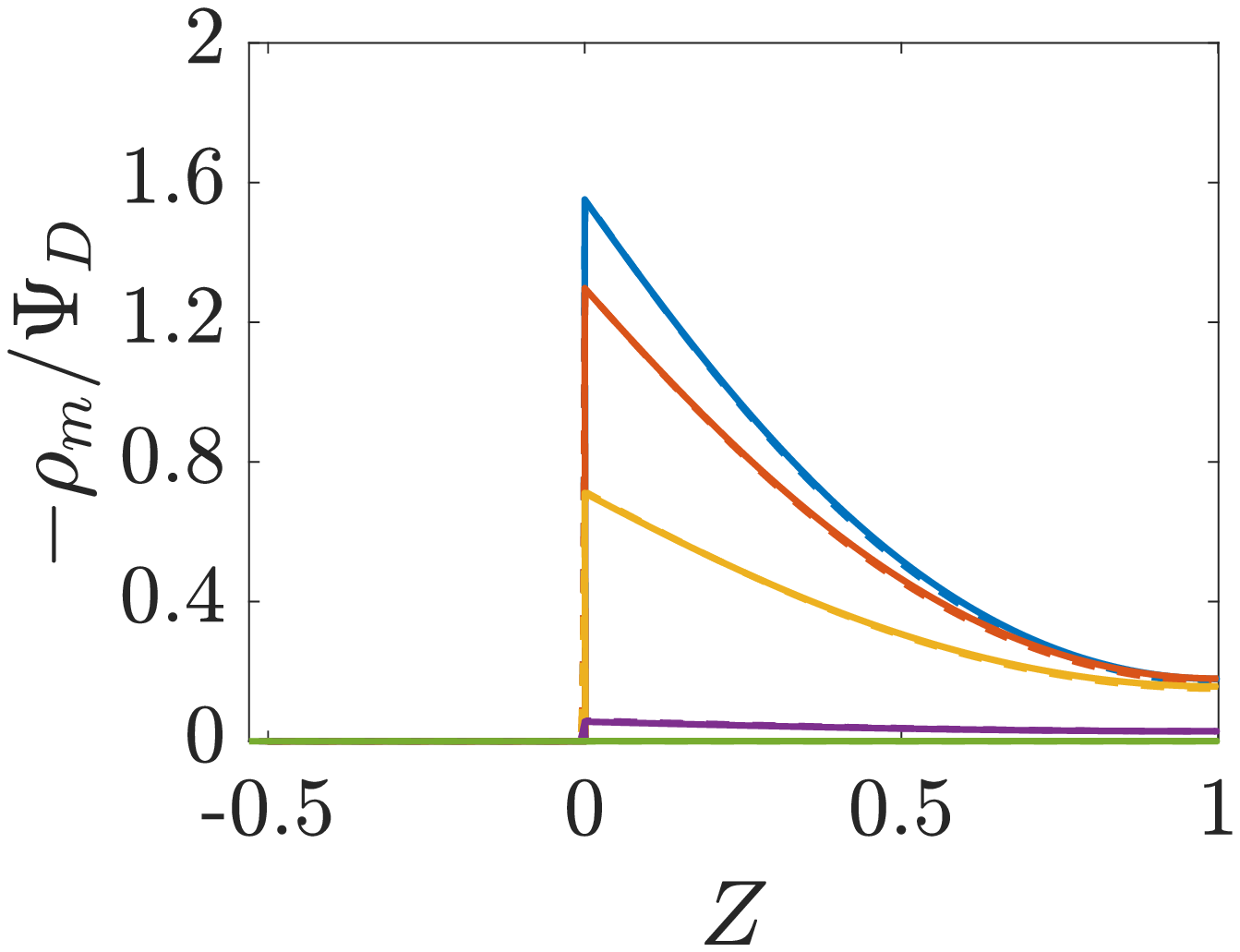}
    \end{subfigure}
    \begin{subfigure}{0.25\textwidth}
        \caption*{\rm{c)}}
        \includegraphics[scale=0.32]{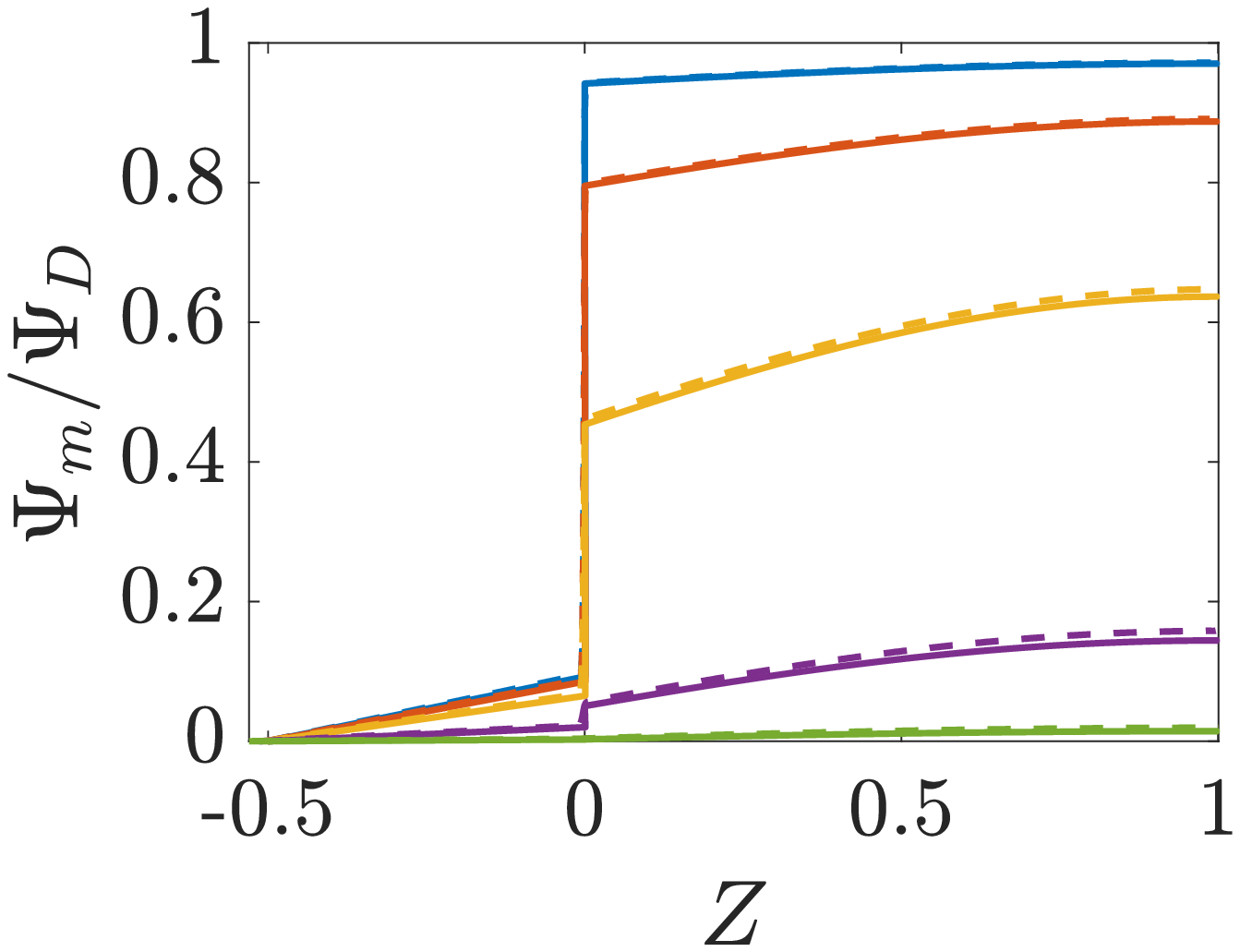}
    \end{subfigure}%
    \begin{subfigure}{0.25\textwidth}
        \caption*{\rm{d)}}
        \includegraphics[scale=0.32]{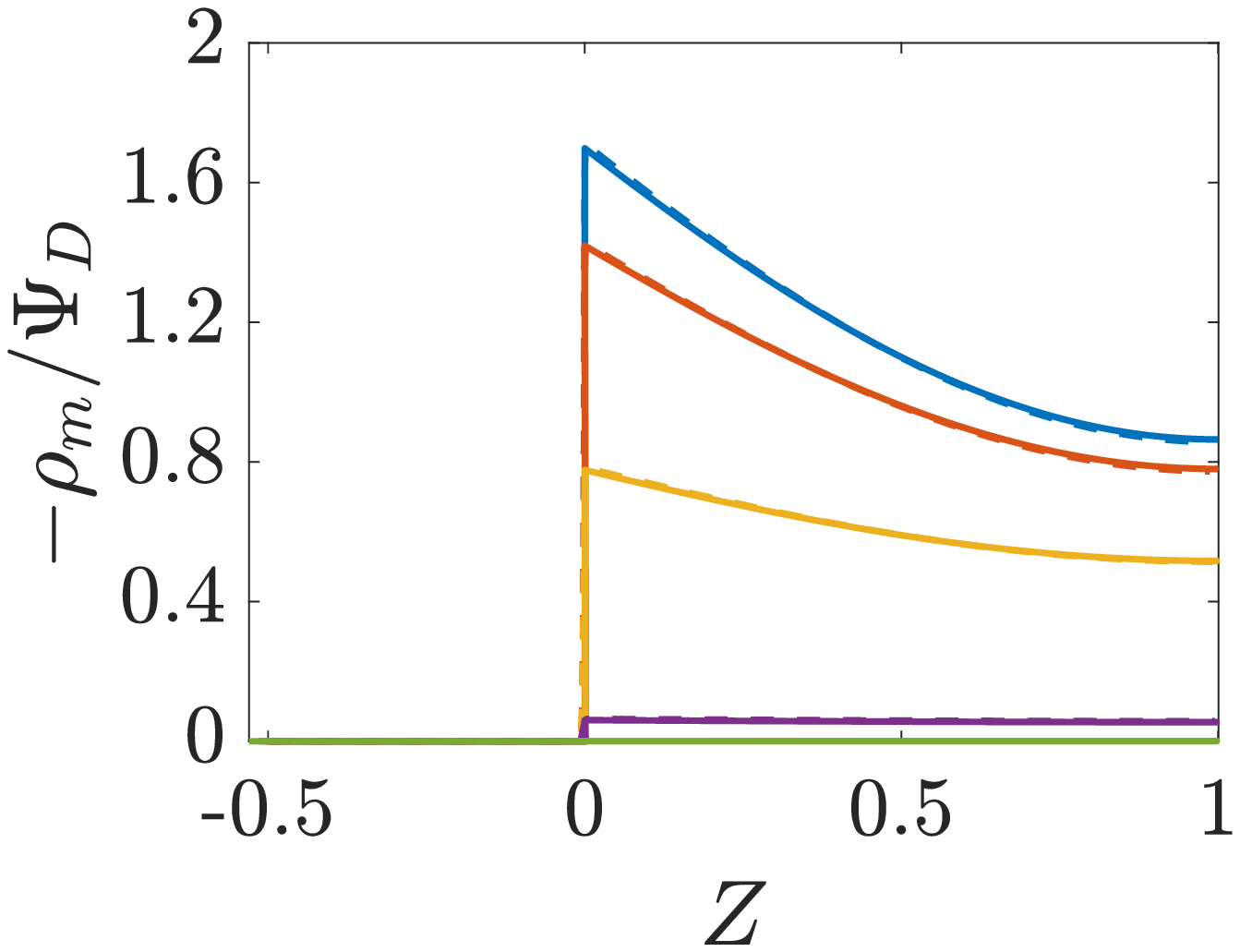}
    \end{subfigure}
    \caption{Centerline potential and charge as a function of the axial coordinate for different $\frac{a_p}{\lambda}$. a) and b) for $\tau=0.15$. c) and d) for $\tau=0.42$. $\psi_D=0.4$ and $\mathrm{Bi}=8$ for all plots. Smaller $\frac{a_p}{\lambda}$ results in an increased centerline potential and charge density, and a larger potential jump at the mouth of the pore. The solid lines are predictions from the perturbation expansion model and the dashed lines are results from DNS.}
    \label{fig:phim}
\end{figure}

In this section, we analyze the analytical predictions of the electrical potential and charge density profiles. We also validate our analytical predictions with the results from DNS.
\subsection{Axial Dependence of Potential and Charge Profiles}
The centerline potential, given by Eq. (\ref{Eq: final_sol}) with $\Psi_m=\phi+\frac{\Psi_D}{I_0(\frac{a_p}{\lambda})}$, and the charge density distribution, given by Eq. (\ref{Eq: rhombessel}), are shown in Fig. \ref{fig:phim}. We note that we obtain excellent agreement between the analytical results and DNS for both $\Psi_m$ and $\rho_m$.

Figs. \ref{fig:phim}b and \ref{fig:phim}d show that the centerline charge density increases monotonically as $\frac{a_p}{\lambda}$ is reduced, vanishing in the thin-double-layer regime, and approaching twice the applied potential for overlapping double layers. At steady state, $\phi\to 0$, $\Psi_m\to \frac{\Psi_D}{I_0(\frac{a_p}{\lambda})}$, and $\rho_m\to -\frac{2\Psi_D}{I_0(\frac{a_p}{\lambda})}$; see Eq. (\ref{Eq: rhombessel}). The increase in centerline charge induces a larger potential change at the SDL-pore { transition} in order to balance diffusion and electromigration in { this} region.

\subsection{Radial Dependence of Potential and Charge Profiles}

\begin{figure}[h!]
    \centering
    \begin{subfigure}{0.25\textwidth}
        \caption*{\rm{a)}}
        \includegraphics[scale=0.32]{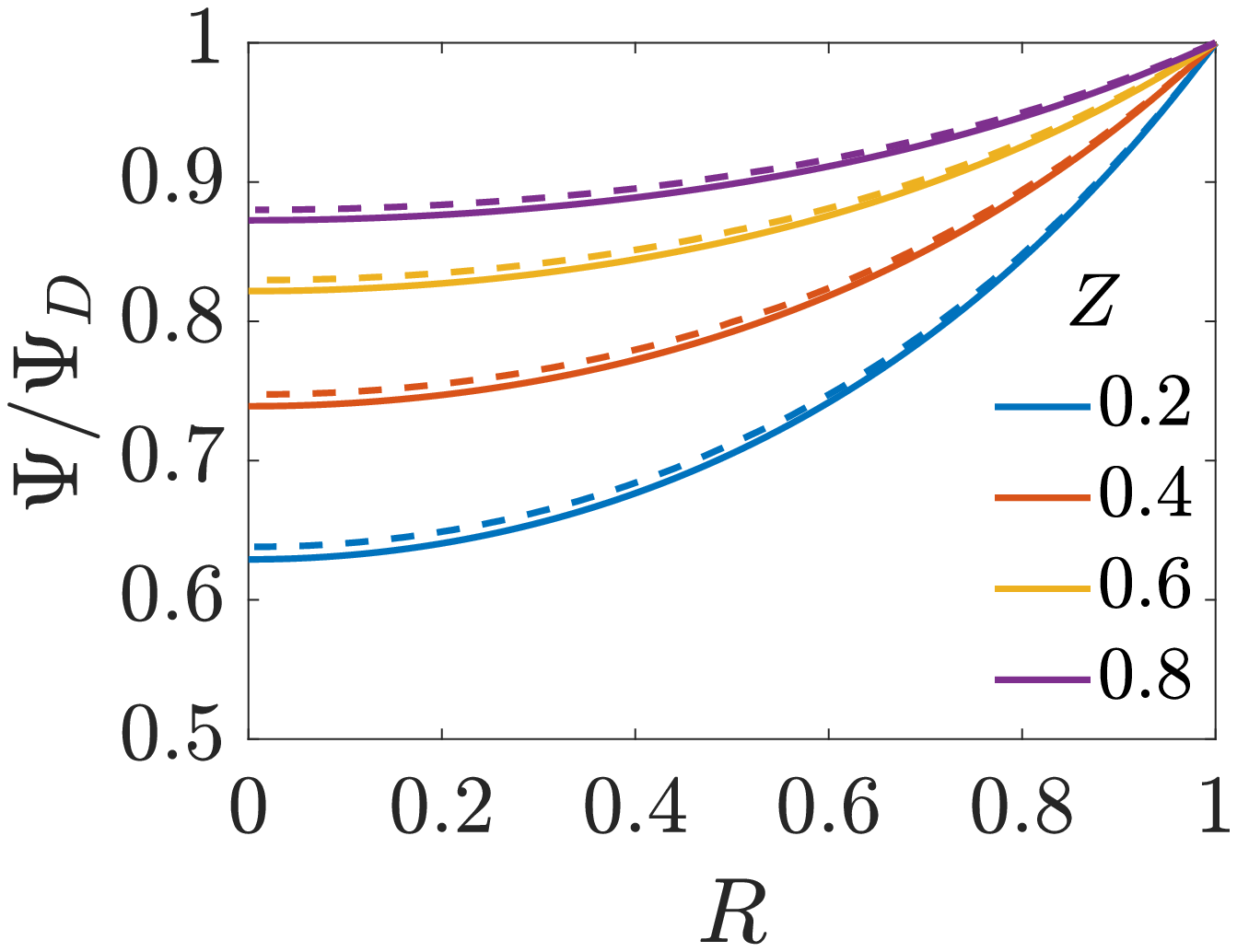}
    \end{subfigure}%
    \begin{subfigure}{0.25\textwidth}
        \caption*{\rm{b)}}
        \includegraphics[scale=0.32]{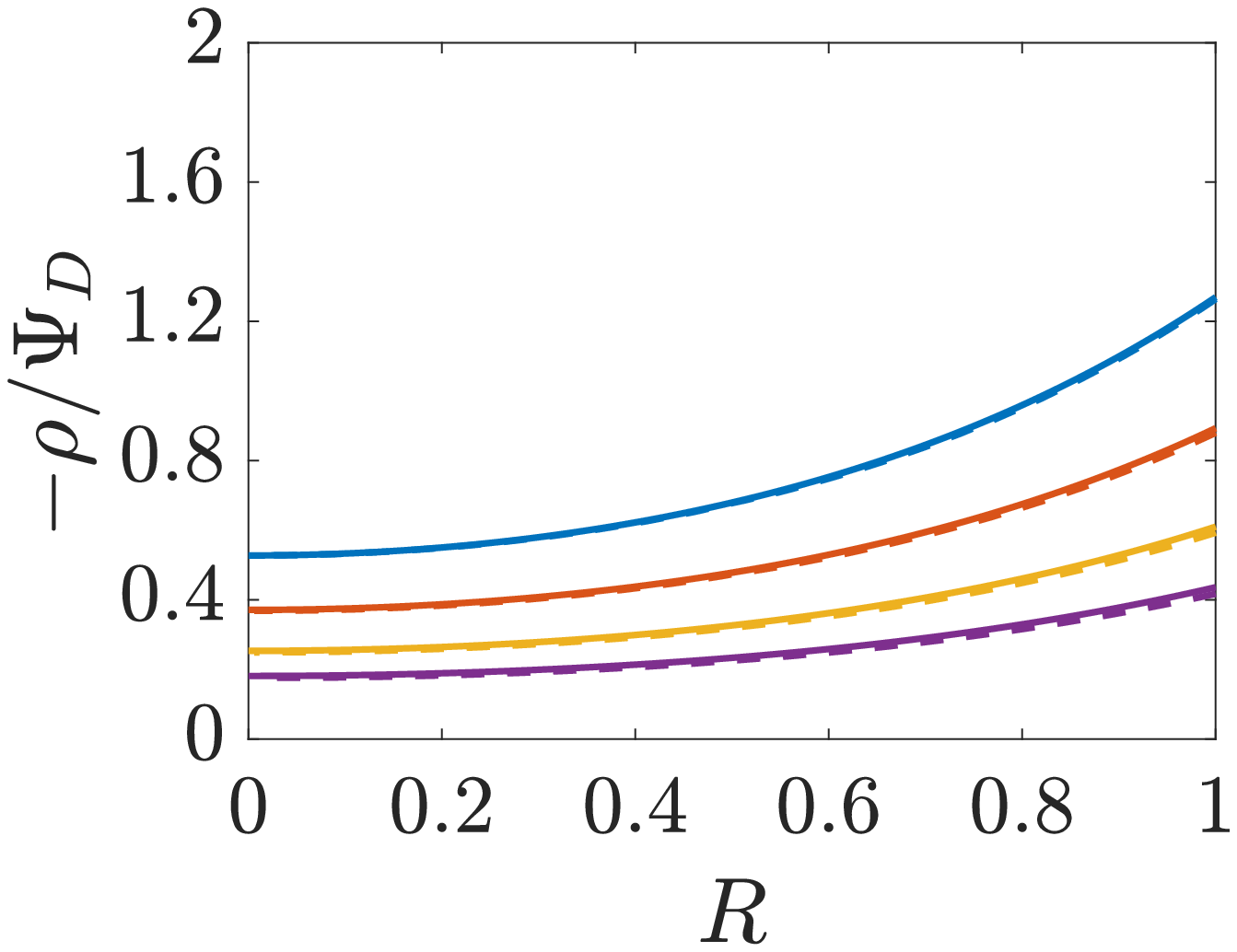}
    \end{subfigure}
    \begin{subfigure}{0.25\textwidth}
        \caption*{\rm{c)}}
        \includegraphics[scale=0.32]{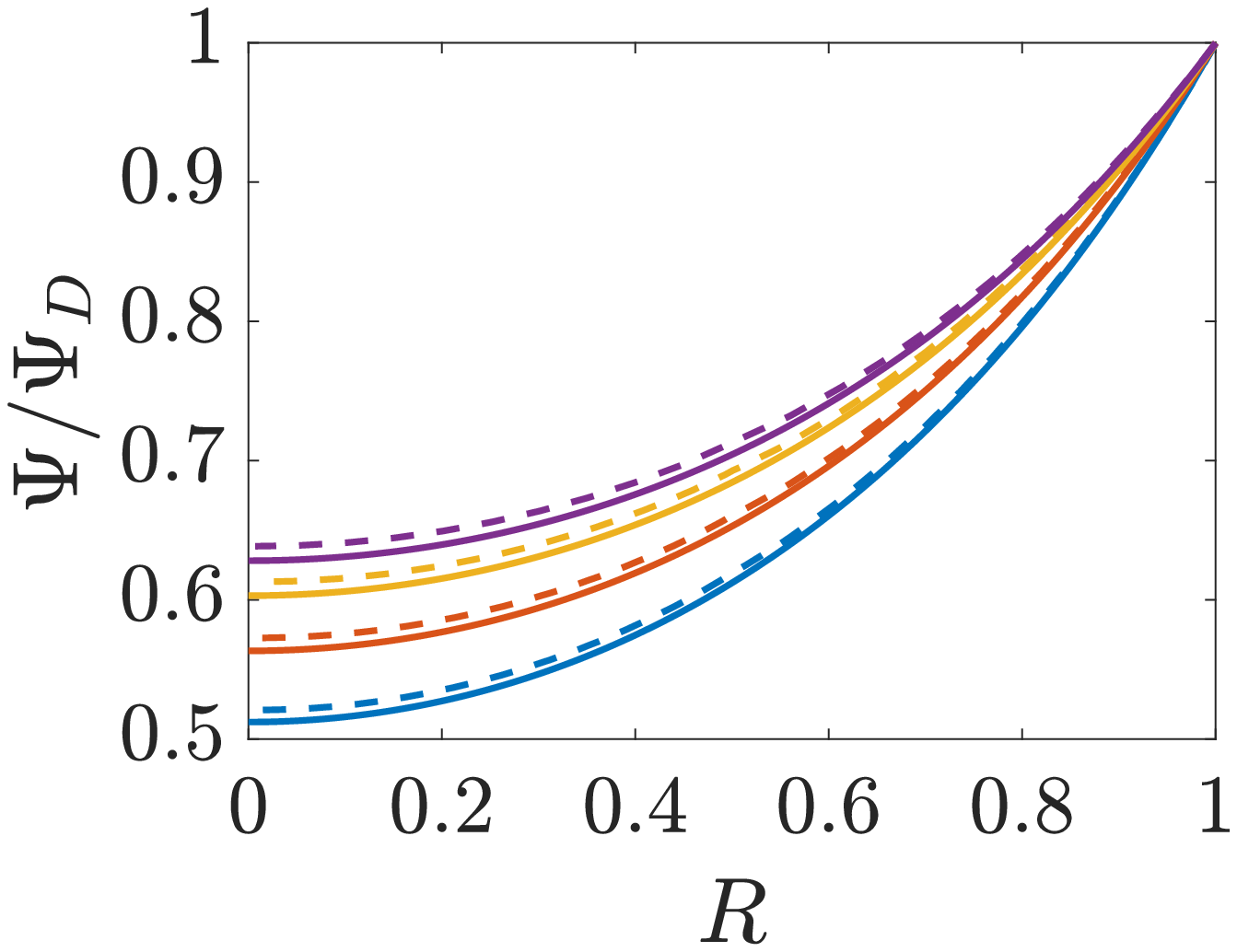}
    \end{subfigure}%
    \begin{subfigure}{0.25\textwidth}
        \caption*{\rm{d)}}
        \includegraphics[scale=0.32]{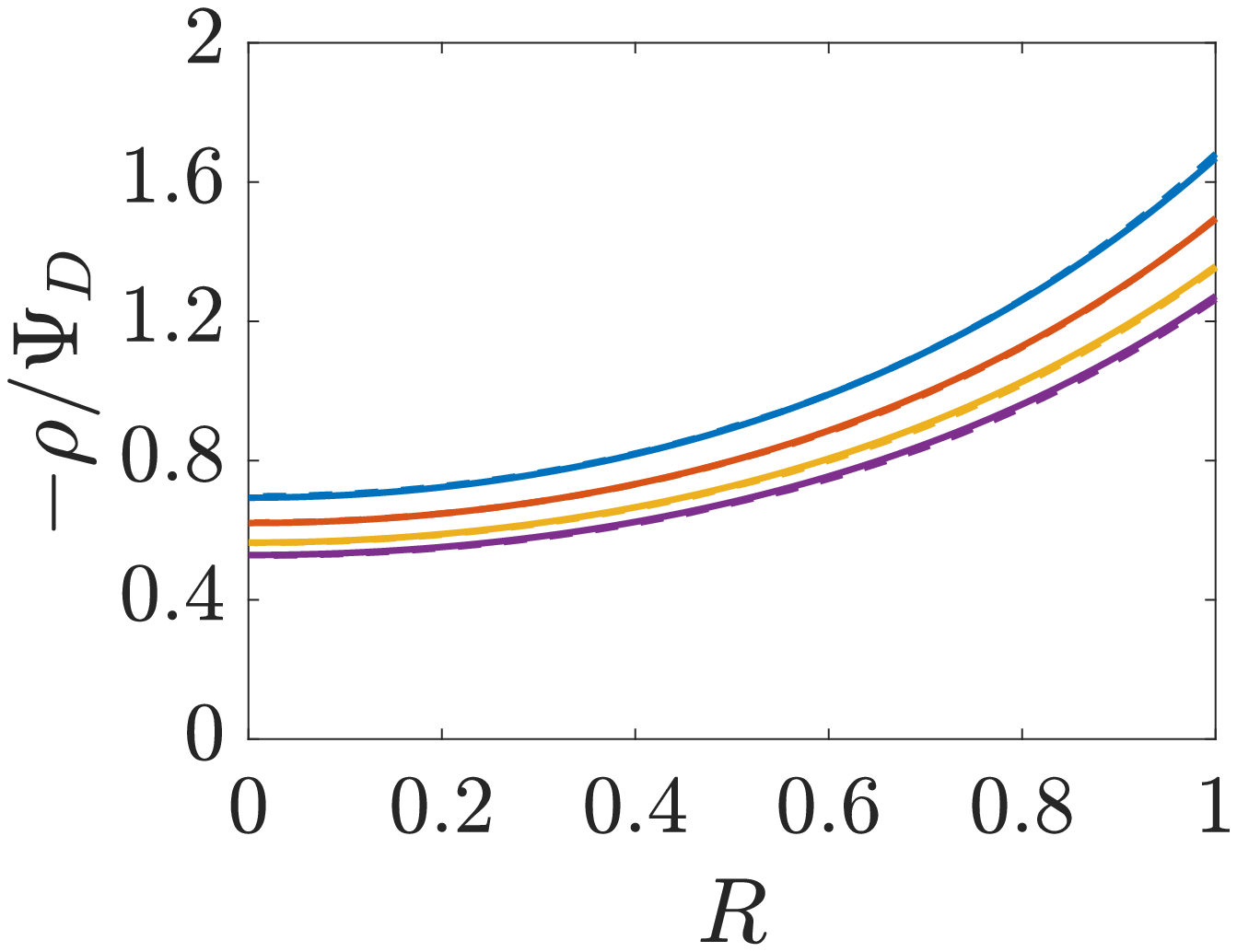}
    \end{subfigure}
    \caption{Potential and charge over the applied potential as a function of the radial coordinate for different axial locations. a) and b) for $\tau=0.15$. c) and d) for $\tau=0.42$. $\psi_D=0.4$, $a_p/\lambda=2$ and $\mathrm{Bi}
    =8$ for all plots. The solid lines are predictions from the perturbation expansion model and the dashed lines are results from DNS.}
    \label{fig:phi}
\end{figure}

The radial dependence of the pore potential and charge distribution, given by Eqs. (\ref{Eq: psival}) and (\ref{Eq: rhoval}), are shown in Fig. \ref{fig:phi} for $\frac{a_p}{\lambda}=2$. The charge distribution propagates gradually along the $Z$-direction, being larger near the mouth of the pore. The potential and charge distributions are related in the radial direction in such a way that the net radial flux vanishes for all times. 
{Figs. \ref{fig:phi}a and \ref{fig:phi}b show opposite dependences of charge density and electric potential on the axial coordinate -- $\rho$ decreases and $\Psi$ increases with $Z$. Essentially, the effect of the charge flux in the SDL is to transport charge into the pore, which subsequently diffuses from the mouth to the end of pore. On the other hand, potential is screened, i.e., reduced, by the charge transported from the SDL. Thus, it is lower closer to the mouth of the pore and increases with $Z$.} At steady state, the charge and potential distributions become independent of $Z$. Using Eqs. (\ref{Eq: psival}) and (\ref{Eq: rhoval}), the distributions for potential and charge throughout the pore at steady state are found to be
\begin{equation}
    \Psi\to \Psi_D\dfrac{I_0(\frac{Ra_p}{\lambda})}{I_0(\frac{a_p}{\lambda})}
\end{equation}
and
\begin{equation}
    \rho\to -2\Psi_D\dfrac{I_0(\frac{Ra_p}{\lambda})}{I_0(\frac{a_p}{\lambda})}.
    \label{eq:rhosteady}
\end{equation}

We note that smaller $\frac{a_p}{\lambda}$ implies higher charge density throughout the pore. However, we {reiterate} that this comes at the cost of a higher charging timescale, since $t_c=\frac{2\frac{\lambda}{a_p}I_1(\frac{a_p}{\lambda})\ell_p^2}{I_0(\frac{a_p}{\lambda})D_p}$.

\subsection{Dependence of Charge Flux on Relative Pore Size}

The centerline potential gives an important physical descriptor of the charging process in the dimensionless charge flux. \par
\begin{figure}[h!]
    \centering
    \begin{subfigure}{0.5\textwidth}
        \centering
        \includegraphics[scale=0.4]{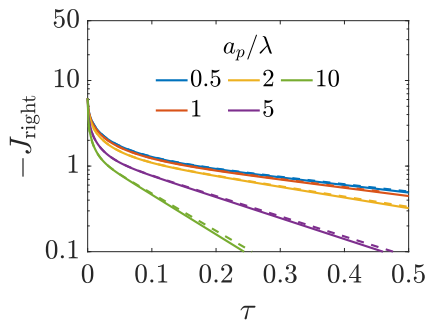}
        \label{}
    \end{subfigure}
    \caption{Dimensionless charge flux at the mouth of the pore ($Z=0^+$) vs. time with $\psi_D$=0.4 and $\mathrm{Bi}=8$. Charge flux decreases monotonically as a function of relative pore size, with exponential decays at late times. The solid lines are predictions from the perturbation expansion model and the dashed lines are results from DNS.}
    \label{fig: current}
\end{figure}

{} Eqs. (\ref{Eq: dim_Ns}), (\ref{Eq: Psi1Psi2}), and (\ref{Eq: final_sol}) can be utilized to show that
\begin{equation}
    J_\mathrm{right}={ -}4\mathrm{Bi}\Psi_D\sum_{n=1}^\infty\dfrac{\sin 2\kappa_n}{2\kappa_n+\sin 2\kappa_n}\exp(-\kappa_n^2T).
    \label{Eq: chargeflux}
\end{equation}
The charge flux comes out to be initially independent of the relative pore size since double layers have not yet formed in the initial state, and the SDL is subjected to the potential gradient imposed by the applied potential; see Eq. (\ref{eq: ics}). The effect of $\frac{a_p}{\lambda}$ is thus to control the charging timescale. In fact, recall that $T= \frac{ I_0 \left( \frac{a_p}{\lambda} \right)}{\frac{2 \lambda}{a_p} I_1 \left( \frac{a_p}{\lambda} \right) } \tau$ and note that $J_\mathrm{right}=J_\mathrm{right}(T)$, alternatively demonstrating that narrower pores take longer to charge. This is illustrated in Fig. \ref{fig: current}, which shows exponential-like behavior at large times, with steeper descents for larger $\frac{a_p}{\lambda}$. 

{ \subsection{Dependence of Charge Flux on the Biot Number}

The ratio of diffusion coefficients in the static diffusion layer and in the pore, $\frac{D_s}{D_p}$, may not be unity. This might be due to different diffusion mechanisms in the SDL and the pore, i.e., bulk diffusion in the former vs. Knudsen diffusion in the latter due to its narrow size \cite{rawlings2002chemical}. The influence of this ratio is encompassed in the Biot number, $\mathrm{Bi}=\frac{A_sD_s\ell_p}{A_pD_p\ell_s}$, which in this case is a ratio of charge transport resistance in the pore vs. in the SDL. Fig. 7 illustrates the dependence of pore charging on the Biot number. It shows that the higher the Biot number, the lower the resistance to charge transfer in the SDL, producing an intuitive effect: enhancement of charge flux for short-times (e.g., owing to higher SDL diffusivity), but also quicker saturation of the pore charge storage. Thus, the order of the curves for different Biot numbers swap over time. We also find that the long-time charging is given by the dominant {timescale} $\tau_c/\kappa_1^2$, where $\kappa_1$ is the first eigenvalue satisfying the characteristic equation $\kappa_n\tan\kappa_n=\mathrm{Bi}$. This is illustrated by the good agreement between the charge fluxes and their approximation by the first mode in the Fourier series of Eq. (\ref{Eq: chargeflux}), represented by dash-dotted lines. It should be noted that net charge storage is not influenced by $\mathrm{Bi}$, as will be shown in Sec. 5.2.\\
\begin{figure}[h!]
    \renewcommand\thefigure{\arabic{figure}}
    \centering
    \includegraphics[scale=0.4]{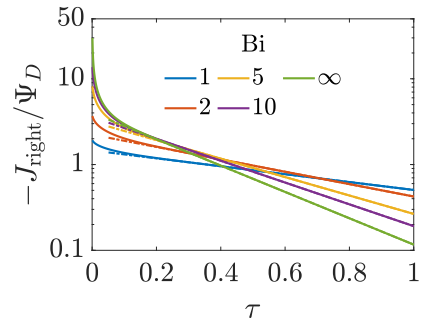}
    \caption{{Influence of the Biot number on the dependence of charge flux with time. Higher Biot implies faster diffusion in the SDL, with an enhancement of the initial charge flux for short-times, but also quicker saturation of the pore charge storage. Continuous lines represent analytical results and the dash-dotted lines show the first mode of the Fourier series in Eq. (30). Relative pore size: $\frac{a_p}{\lambda}=2$. }
    \label{fig:JBi}
    }
\end{figure}
}

\subsection{Validity for Higher Potentials}

{For higher applied potentials, the full PNP equations assuming radial equilibrium and relaxing the constraint of constant salt follow from Eqs. (\ref{Eq: dim_pnp}) as
\begin{equation}
    \dfrac{\partial \rho}{\partial \tau}=\dfrac{\partial^2\rho}{\partial Z^2}+\dfrac{\partial}{\partial Z}\left(s\dfrac{\partial\Psi}{\partial Z}\right),
\end{equation}
\begin{equation}
    \dfrac{\partial s}{\partial \tau}=\dfrac{\partial^2 s}{\partial Z^2}+\dfrac{\partial}{\partial Z}\left(\rho\dfrac{\partial\Psi}{\partial Z}\right).
\end{equation}
The main simplification that comes from the linear regime $\Psi_D\ll 1$ is the neglect of the term $\rho\frac{\partial\Psi}{\partial Z}$, whence the salt transport equation resumes to a transient diffusion equation. Given the initial and boundary conditions for salt, the trivial solution $s(R,Z,\tau)=2$ holds, i.e., we can assume salt to be constant. For higher potentials, though, that approximation ceases to be valid and the variation of salt density in the domain influences the strength of electromigration of charge.}

\begin{figure}[h!]
    \centering
    \begin{subfigure}{0.25\textwidth}
        \centering
        \caption*{\rm{a)}}
        \includegraphics[scale=0.32]{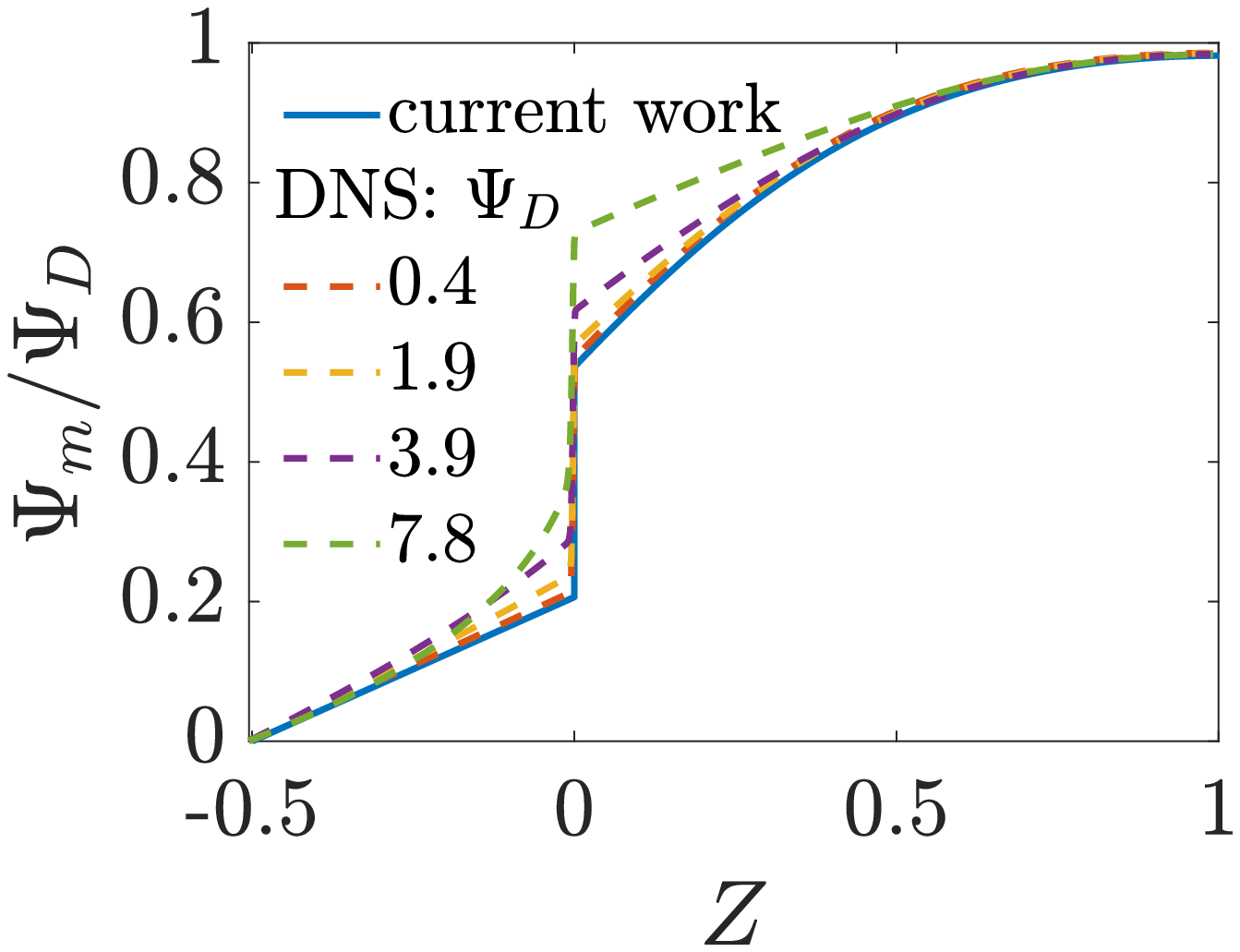}
        \label{pota}
    \end{subfigure}%
    \begin{subfigure}{0.25\textwidth}
        \centering
        \caption*{\rm{b)}}
        \includegraphics[scale=0.32]{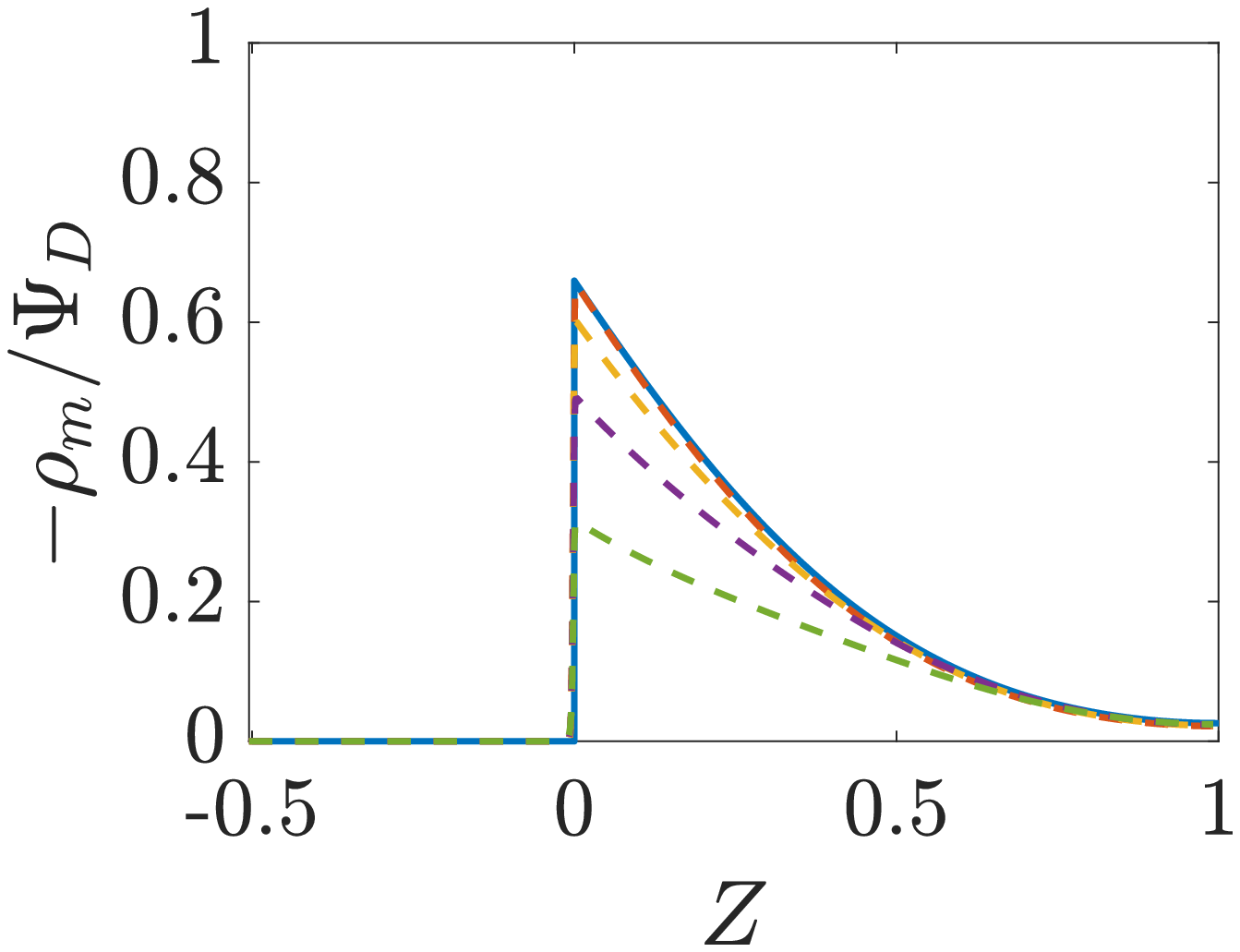}
        \label{potb}
    \end{subfigure}
    \begin{subfigure}{0.25\textwidth}
        \centering
        \caption*{\rm{c)}}
        \includegraphics[scale=0.32]{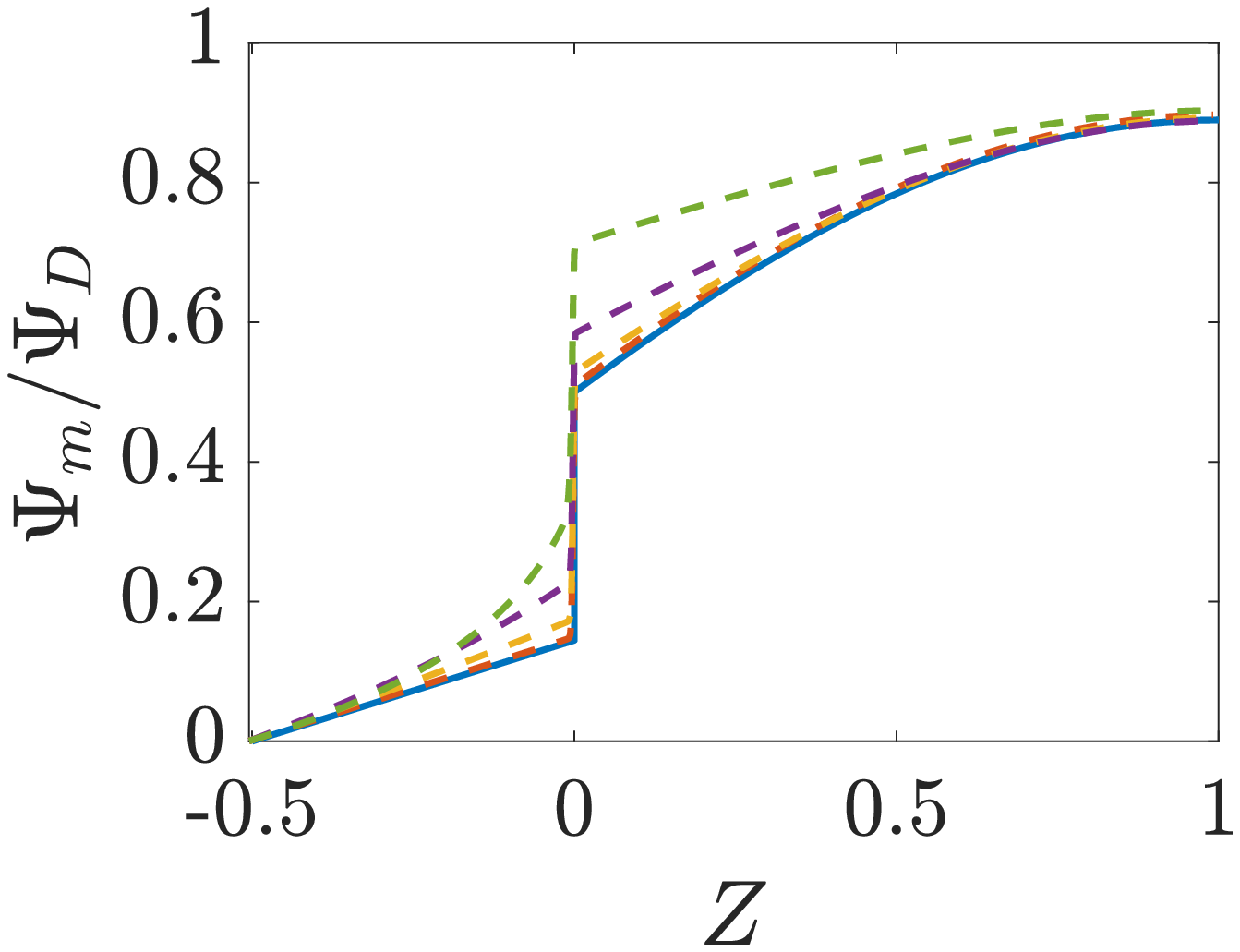}
        \label{potc}
    \end{subfigure}%
    \begin{subfigure}{0.25\textwidth}
        \centering
        \caption*{\rm{d)}}
        \includegraphics[scale=0.32]{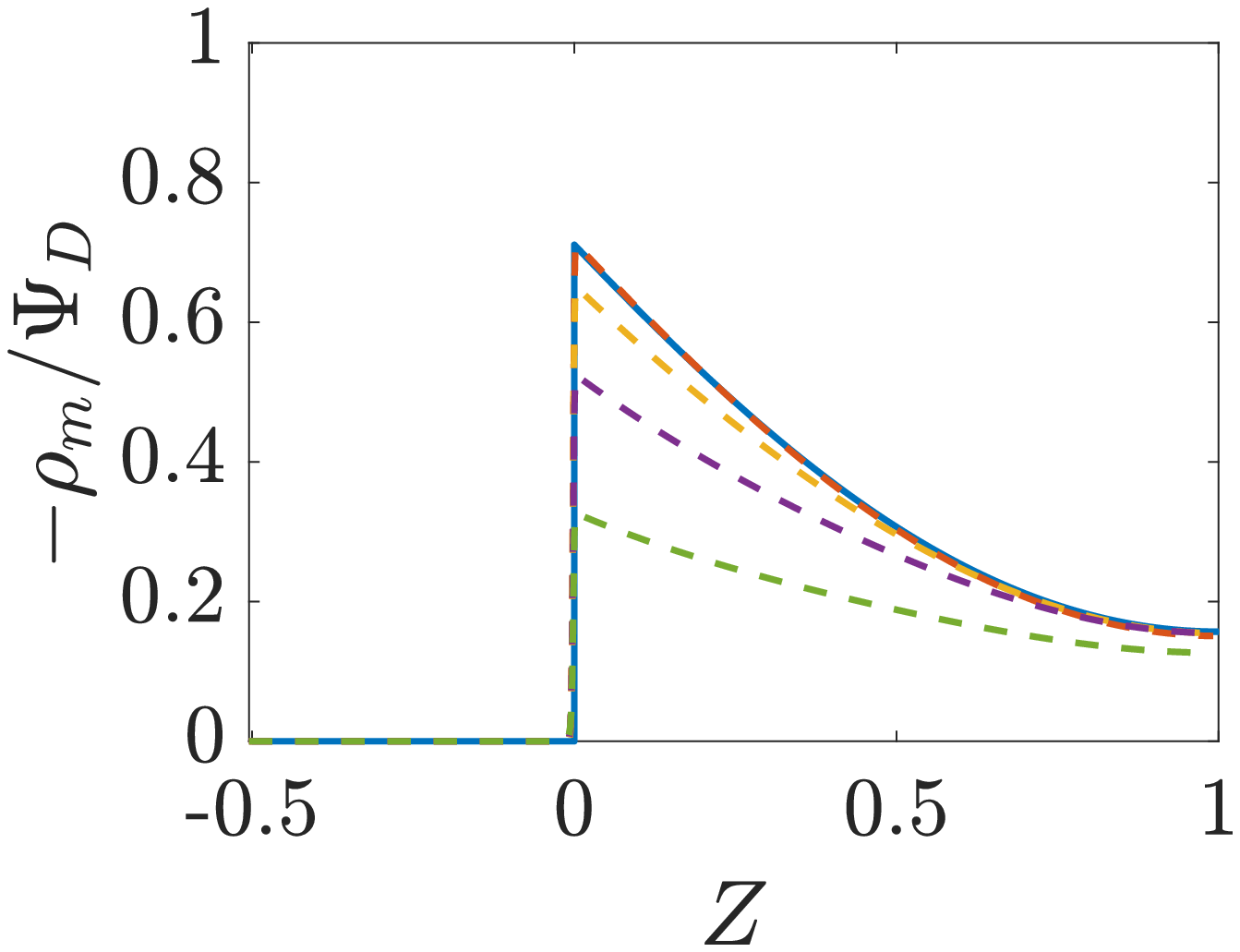}
        \label{potd}
    \end{subfigure}
    \begin{subfigure}{0.5\textwidth}
        \centering
        \caption*{\rm{e)}}
        \includegraphics[scale=0.32]{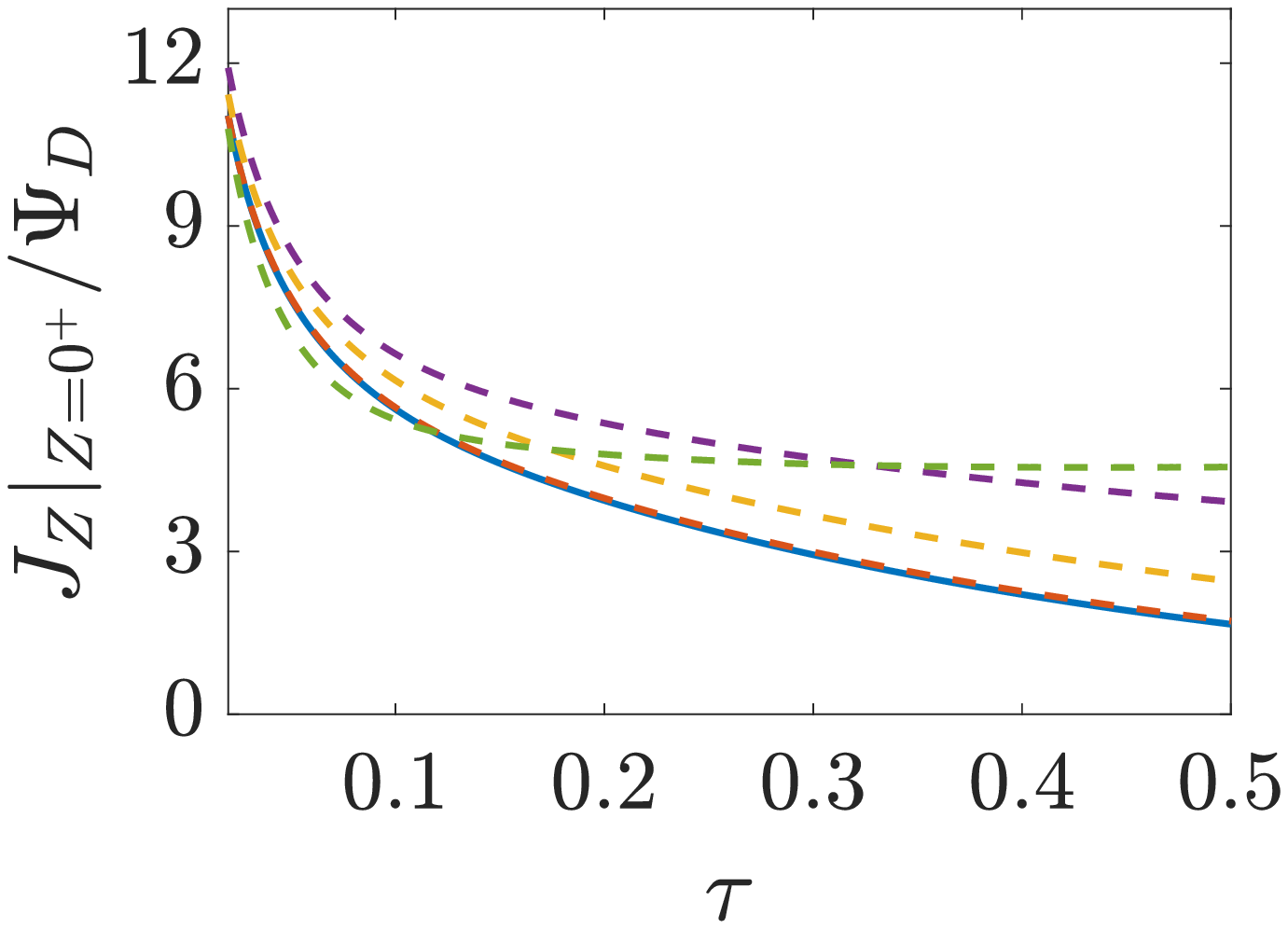}
        \label{pote}
    \end{subfigure}
    \caption{Validity of analytical results. Comparison of the proposed model (continuous lines) with DNS results (dashed lines) for different applied potentials. a) and b) $\tau=0.07$, c) and d) $\tau=0.15$.
    e) Time dependence of the charge flux for $a_p/\lambda=2$. Good agreement is obtained up to twice the thermal potential ($\approx$ 50mV at room temperature). The current shows good agreement for early times even for moderate potentials, but salt corrections to charge electromigration must be taken into account for long times.}
    \label{fig:psidvalidation}
\end{figure}

In order to assess the validity of our analysis, beyond which the neglect of higher-order corrections in $\Psi_D$ becomes unwarranted, we compare our results to the DNS for different applied potentials in Fig. \ref{fig:psidvalidation}. We obtain good agreement between potentials and charge profiles until the applied potential reaches approximately twice the thermal potential, or approximately 50 mV at room temperature. Above that threshold, a non-linear potential response in the SDL is observed in the DNS, and thus the first-order corrections underpredict the potential in the pore and overpredict the charge stored. Indeed, Fig. \ref{fig:psidvalidation}e shows that the charge flux predictions agree with DNS results for early times ($\tau<0.1$), even for moderately high applied voltages, with an error of 8\% for 200 mV (but of 15\% for 100 mV). However, for late times, { non-linear terms become important, introducing electromigration of salt which can { affect} the transport of charge.}

\section{Analysis of Capacitance}
Several studies on electrode charging invoke transmission line representations, theoretically developed for thin double layers \cite{de1963porous,de1964porous,janssen2021transmission}, to address electric potential and capacitance in these systems \cite{black2010pore,kaus2010modelling,kowal2011detailed,madabattula2018insights}. In this section, we demonstrate that a similar transmission line model can be constructed for arbitrary pore sizes. However, a time-dependent {transition-region resistor} makes it challenging to compare it directly with experiments. We also derive an effective capacitance for arbitrary pore sizes from the steady-state charge density profiles.

\subsection{Transmission Line Circuit}\label{sec: tl}

\par{}First, we derive the value of conductivity for arbitrary $\frac{a_p}{\lambda}$. Briefly restoring dimensions and utilizing Eqs. (\ref{Eq: dim_Ns}), (\ref{Eq: psival}) and (\ref{Eq: rhoval}), we find that axial charge flux $J_z$ inside the pore is given as
{
\begin{equation}
    J_z = - \frac{2 D_p e^2 c_0}{k_B T} \frac{I_0 \left( \frac{a_p}{\lambda} \right)}{I_0 \left( \frac{a_p}{\lambda} \right) - 1}\frac{\partial \psi_m}{\partial z}=-\tilde{\sigma}_p\frac{\partial \psi_m}{\partial z}, 
\end{equation}}
such that dimensional pore conductivity is given by { ${\tilde{\sigma}_p = \frac{2 D_p e^2 c_0}{k_B T} \frac{I_0 \left( \frac{a_p}{\lambda} \right)}{I_0 \left( \frac{a_p}{\lambda} \right) - 1}}$}. Note that the conductivity for $\frac{a_p}{\lambda} \gg 1$ is only based on electromigrative charge flux. In contrast, the conductivity for $\frac{a_p}{\lambda} \ll 1$ is enhanced because both diffusive and electromigrative fluxes contribute to current; see Fig. (\ref{fig:fluxvec}).
\par{} Next, we derive the dimensional capacitance per unit surface area, $\tilde{C}_p$. We note that $2 \pi a_p \tilde{C}_p (\psi_m -\psi_D)$ should yield the total charge stored per unit axial length. By utilizing Eq. (\ref{Eq: rhoval}) and integrating along the radial direction, it is straightforward to show that  $\tilde{C}_p = \frac{\varepsilon}{\lambda} \frac{I_1 \left( \frac{a_p}{\lambda} \right) }{I_0 \left( \frac{a_p}{\lambda} \right) - 1} $. In the thin-double-layer limit $\frac{a_p}{\lambda} \gg 1$, we recover the usual expression that $\tilde{C}_p \rightarrow \frac{\varepsilon}{\lambda}$. In contrast, in the thick-double-layer limit $\frac{a_p}{\lambda} \gg 1$, we obtain $\tilde{C}_p \rightarrow \frac{2 \varepsilon}{a_p}$. This result is intuitive since it demonstrates that the length scale of capacitance in the overlapping-double-layer limit is controlled by the pore size. Based on these expressions, we can also recover $t_c = \frac{2 \pi a_p \tilde{C}_p}{\pi a_p^2 \tilde{\sigma}_p} \ell_p^2 = \frac{\frac{2 \lambda}{a_p} I_1 \left( \frac{a_p}{\lambda} \right) }{ I_0 \left( \frac{a_p}{\lambda} \right)} \frac{\ell_p^2}{D_p}$, consistent with Eq. (\ref{eq: tc}).
\par
\begin{figure*}[t]
    \centering
    \valign{#\cr
      \hsize=0.65\columnwidth
        \begin{subfigure}{0.65\textwidth}
            \centering
            \caption*{\rm{a)}}  \includegraphics[scale=1.0]{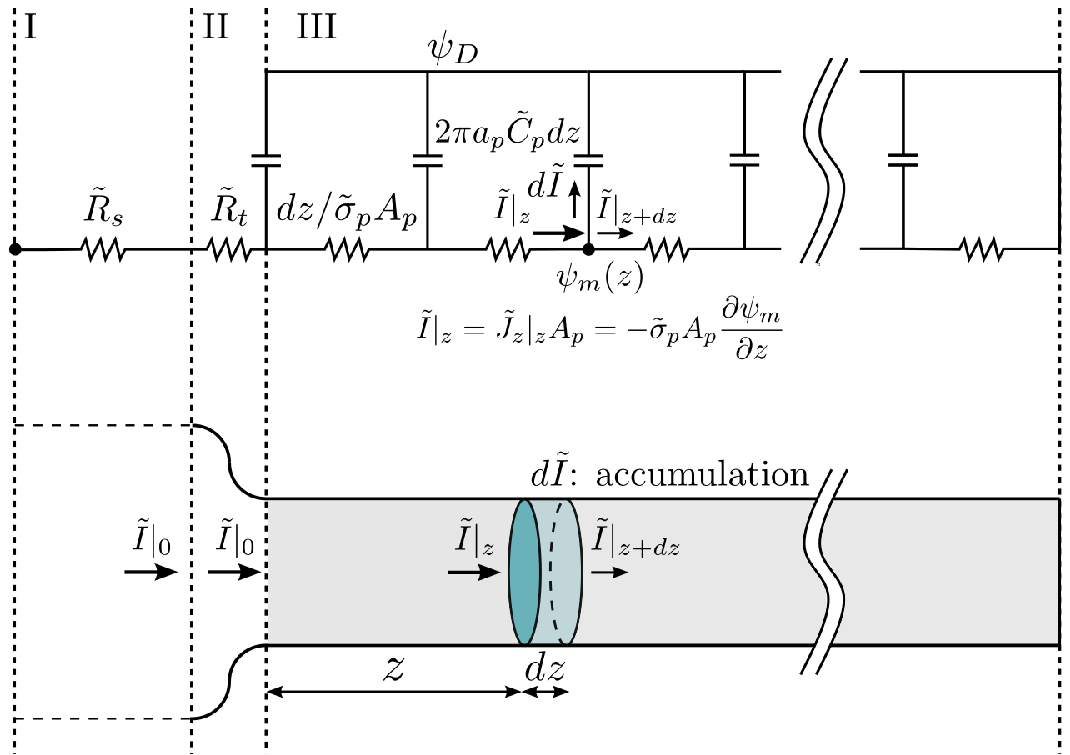}
            \label{fig:circuita}
        \end{subfigure}\cr\noalign{\hfill}
      \hsize=0.75\columnwidth
        \begin{subfigure}{0.25\textwidth}
            \centering
            \caption*{\rm{b)}}
            \includegraphics[scale=0.33]{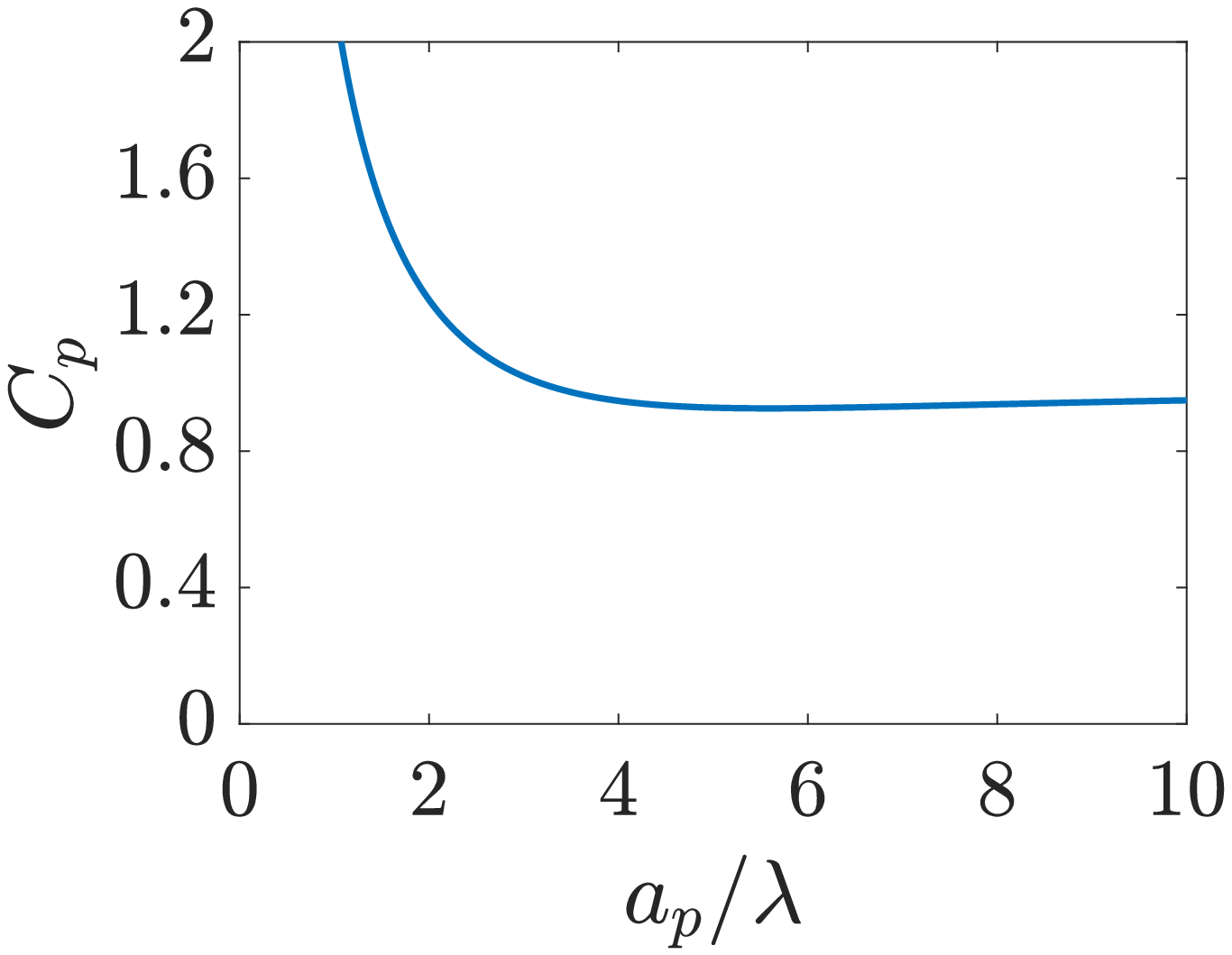}
            \label{fig:circuitb}
        \end{subfigure}\vfill
        \begin{subfigure}{0.25\textwidth}
            \centering
            \caption*{\rm{c)}}
            \includegraphics[scale=0.33]{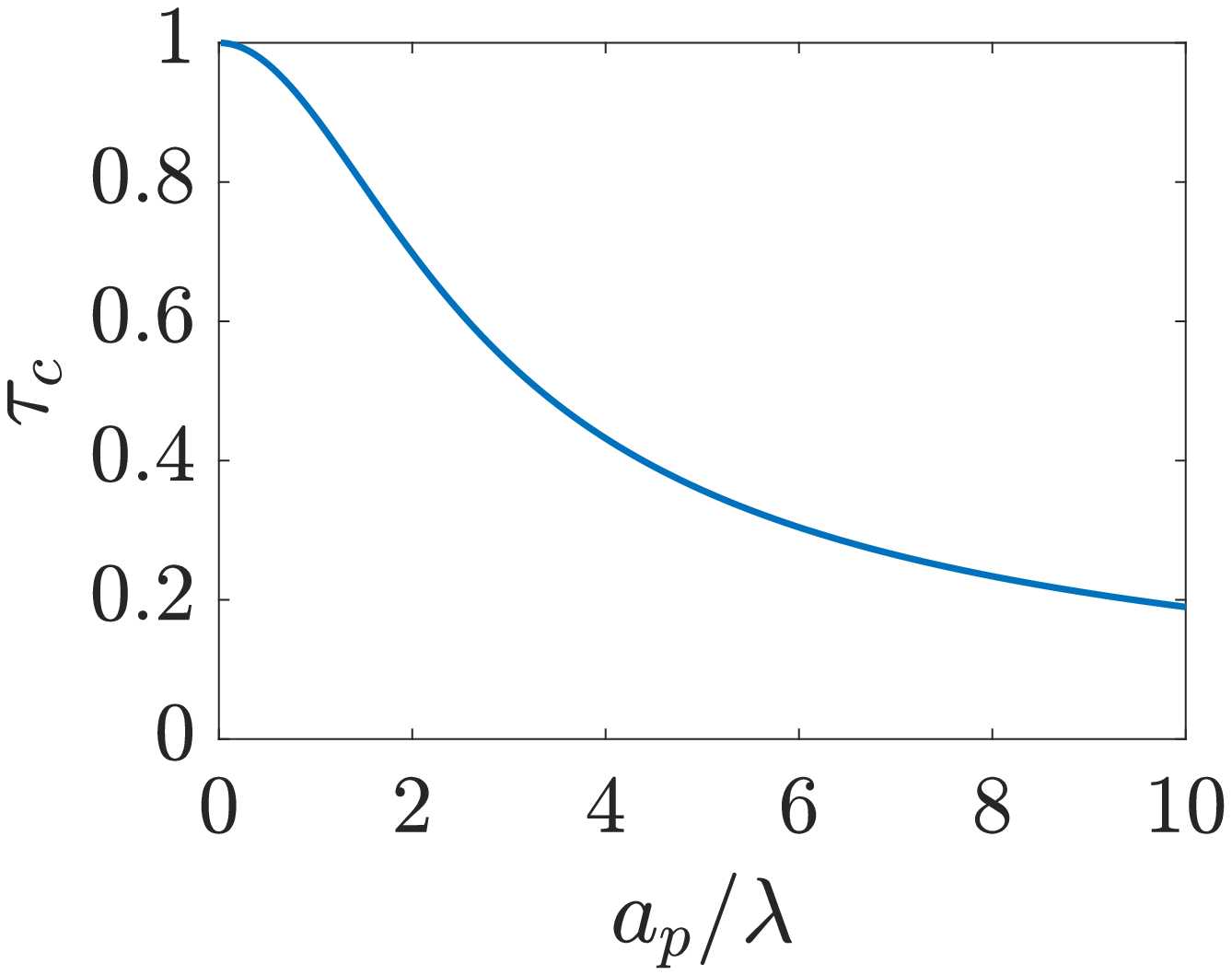}
            \label{fig:circuitc}
        \end{subfigure}\cr
    }
    \caption{{a) Top panel: Transmission line circuit schematic for arbitrary pore size. I: SDL, II: transition region, III: pore. SDL and transition resistances given by Eqs. (\ref{eq:Rsdim}) and (\ref{eq:Rtdim}). Bottom panel: equivalent representation of the charge balances in an infinitesimal volume in a pore. b) dimensionless pore capacitance per unit surface area, ${C}_p=\tilde{C_p}\frac{\lambda}{\varepsilon}$, c) dimensionless charging time, $\tau_c=t_c\frac{D_p}{\ell_p^2}$. With the presence of a transition-region resistor, the transmission line model becomes more intricate and loses some applicability.}}
    \label{fig:circuit}
\end{figure*}

{} The linear relation between average diffusive and electromigrative axial fluxes allowed us to determine expressions for pore capacitance and resistance that satisfy Ohm's law and the definition of a capacitive element. However, the key distinguishing feature from the classical thin double layer analysis \cite{de1963porous,de1964porous} is the inclusion of potential change across the transition region as derived in Eq. (\ref{Eq: Psi1Psi2}), which comes from the charge flux matching. In order to account for this change in potential, we add a { resistor} representing the interface, as shown in Fig. {\ref{fig:circuit}a}. { Its dimensional resistance is determined from Ohm's law,
    \begin{equation}
        \tilde{R}_\mathrm{t}=\dfrac{2\lambda^2\ell_p}{\varepsilon D_pA_p}\dfrac{\Psi_\mathrm{left}-\Psi_\mathrm{right}}{J_{\mathrm{right}}}.
    \end{equation}
We derive an expression for {dimensionless transition-region resistance}  $R_\mathrm{t}(\tau)$ in Appendix A along with the traditional expression for SDL resistance, since their final expressions are not utilized for further analysis.}\par{}
{ Figs. \ref{fig:circuit}b and \ref{fig:circuit}c show} the pore capacitance and charging time as per the transmission line model. The sharp increase in pore capacitance per unit surface area for smaller $\frac{a_p}{\lambda}$ may suggest {a blowup of charge flux for overlapping double layers, which is not observed in Fig. \ref{fig: current}. In fact, though pore capacitance increases, pore resistance decreases commensurately, in a way that yields $\tau_c\sim 1$ for overlapping double layers. Even more importantly, pore capacitance of arbitrary $\frac{a_p}{\lambda}$ reported in our manuscript is useful for the transmission circuit analysis that predicts centerline potential, but since it's based on the potential difference $\Psi_m-\Psi_D$, it is not representative of  experimental measures of capacitance, which are based on $\Psi_D$ \cite{zhang2014highly,boota2015graphene}. This is the motivation for the definition of an effective capacitance that we develop in the next section.}

\subsection{Macroscopic Perspective}\label{sec: macro}

\begin{figure}[t]
    \centering
    \includegraphics[scale=0.4]{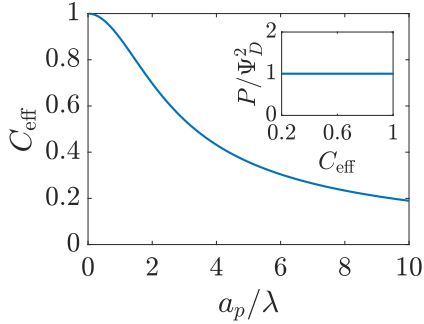}
    \caption{Dimensionless effective volumetric pore capacitance versus relative pore size. Narrower pores account for higher volumetric capacitances. The inset shows power density versus effective volumetric pore capacitance. Gains in capacitance for a change in pore size occur at constant power density.}
    \label{fig:capeff}
\end{figure}

Despite the possibility of representing pore charging by the transmission line circuit described in Sec. \ref{sec: tl}, its time-dependent {transition-region resistor} enters as an additional factor influencing the pore charging performance. {In addition, the definition of pore capacitance based on the potential difference $\Psi_m-\Psi_D$ makes it incompatible with experimental measures}. Therefore, in the interest of facilitating the analysis of pore-size effects on charging performance, our focus lies on a direct comparison via a macroscopic perspective, i.e., from characterizing the total charge stored by the pore at steady state. We employ the definition of effective volumetric capacitance $C_{\mathrm{eff}}$ as the ratio of total charge stored per applied voltage and unit volume. Going back to dimensionless variables and performing a radial integration of the steady-state charge density profile of Eq. (\ref{eq:rhosteady}), it is straightforward to obtain 
\begin{equation}
    C_{\textrm{eff}} =  \frac{  \frac{2 \lambda}{a_p}   I_1 \left( \frac{a_p}{\lambda} \right)}{I_0 \left( \frac{a_p}{\lambda} \right)},
    \label{Eq: ceffvol}
\end{equation}
{where capacitance is scaled by $\frac{\varepsilon}{\lambda^2}$}. Eq. (\ref{Eq: ceffvol}) shows that $C_{\textrm{eff}}=\tau_c$, i.e., dimensionless volumetric capacitance is equal to dimensionless charging time; { see Figs. \ref{fig:circuit}c and \ref{fig:capeff}}. The latter shows that the volumetric capacitance decreases with an increase in relative pore size{, that is, overlapping double layers present optimal energy storage. A similar behavior of energy density increase with width reduction due to a decrease of the electroneutral region has been reported in Refs. \cite{kondrat2012effect} for nanopores, and \cite{varghese2011simulating} for pores ranging from 0.5 to 10 nm. However, in our model,} this comes at a cost of a corresponding increase in the charging time $\tau_c$. From an engineering perspective, this prediction means that an increase in the energy density,  {$E=C_{\textrm{eff}}\Psi_D^2$}, with a change in relative pore size is unaccompanied by changes in power density, {$P=E/\tau_c$; see the inset in Fig. \ref{fig:capeff}}. In fact, this relationship between energy and power density is {consistent with experiments}; see Fig. 1 of Ref. \cite{simon2008materials}. To the best of our knowledge, this interplay between energy and power density and its dependence on pore sizes hasn't been derived previously.
\subsection{Effect of Pore-Size Distribution}
\par
\begin{figure}[h!]
    \centering
    \begin{subfigure}{0.25\textwidth}
        \centering
        \caption*{\rm{a)}}
        \includegraphics[scale=0.31]{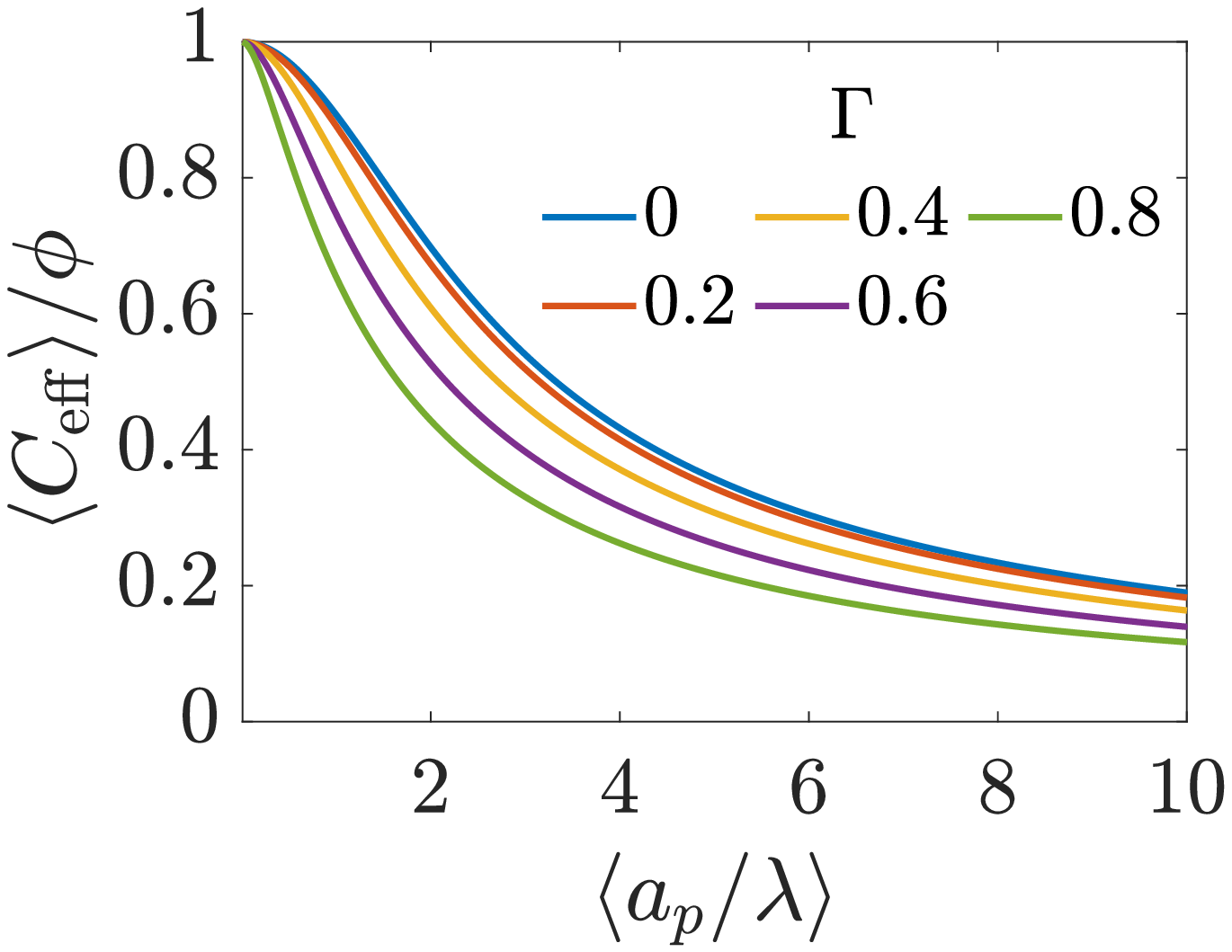}
        \label{fig:pdf}
    \end{subfigure}%
    \begin{subfigure}{0.25\textwidth}
        \centering
        \caption*{\rm{b)}}
        \includegraphics[scale=0.31]{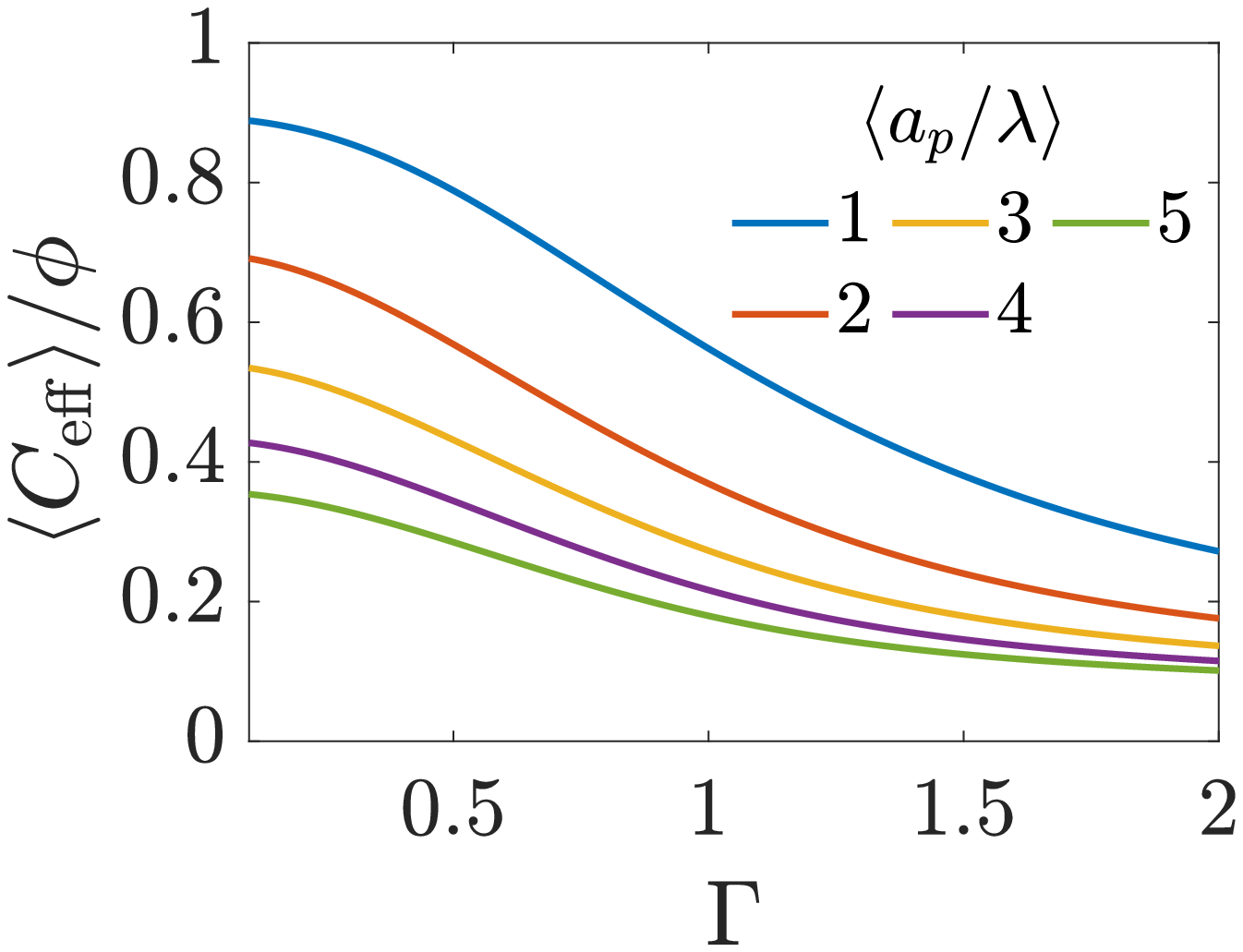}
        \label{fig:cv}
    \end{subfigure}
    \caption{ a) Dimensionless average electrode capacitance per unit volume over the electrode porosity $\langle C_{\mathrm{eff}}\rangle/\phi$ vs. average { relative} pore size $\langle\frac{a_p}{\lambda}\rangle$ for a log-normal size distribution. b) dimensionless averaged electrode capacitance per unit volume vs. polydispersity of the pore-size distribution, {$\Gamma$, i.e., standard deviation of relative pore size over average relative pore size}. An increase in the polydispersity implies a higher frequency of wider pores, which results in a decrease of the electrode capacitance.}
    \label{fig:ceffavg}
\end{figure}

{} An important advantage of our analysis is that it is valid for an arbitrary $\frac{a_p}{\lambda}$, thus it can also be conducted for distributions of pore sizes to study the impact of polydispersity. {Within the limitations of our model, i.e., specificity to non-interacting pores, we propose a simplified model to examine the influence of double-layer thickness over electrode charging.}
To this end, we assume a log-normal probability distribution function of pore sizes, in consonance with {experimentally measured pore-size distributions \cite{jorne1986effect,song1999electrochemical}}, and perform averages of the pore properties described in Sec. \ref{sec: macro} over the pore-size distribution; see Appendix C. In our analysis, we determine the effects of pore-size average and { polydispersity}. Fig. \ref{fig:ceffavg} shows that {distributions with lower averages or polydispersities of relative pore sizes present higher average capacitances} (i.e., the electrode capacitance) due to elevated volumetric capacitance of pores with low relative pore size.
    Though this conclusion is specific to the probability density function employed, the principle of boosting {average} capacitance {with regards to double layer thickness} by increasing the frequency of narrow pores {should hold} in general. { Optimal energy density for monodisperse pore distributions has also been reported in Monte Carlo simulations of nanopores \cite{kondrat2012effect}.} Nevertheless, the one caveat that also follows {from our analysis} is the accompanying increase in electrode charging time. 

\par{} There is an ongoing debate about the experimentally observed behavior of capacitance for sub-nanometer pores. While some studies report an anomalous areal capacitance (i.e., capacitance per unit area) increase in sub-nanometer pores \cite{chmiola2006anomalous,huang2008theoretical, simon2008materials}, other works claim that it is roughly independent of pore size in that regime\cite{centeno2011capacitance,garcia2015constant}, the discrepancy being attributed to inaccuracies in BET isotherm surface area determination for subnanometer pores. The results of our work { seem to support the latter hypothesis}, showing only a mild increase of areal capacitance with pore size for $\frac{a_p}{\lambda}>2$ (corresponding to $a_p>0.6$ nm for 1 M electrolytes at room temperature); see Appendix A for calculations details.  { However,} we acknowledge that our model may fail to capture intricacies of subnanometer pores, such as anomalous capacitance increases due to loss of solvation shells \cite{huang2008theoretical, simon2008materials}. This is expected since the Poisson-Nernst-Planck equations do not take into account finite ion-radius and confinement effects \cite{Kornyshev2007}, which will become crucial in the subnanometer regime.

\section{Conclusion}
\label{sec: conc}

In summary, in this article:
\begin{itemize}
\item A regular perturbation expansion model for double layer charging at arbitrary pore sizes is proposed. The effects of arbitrary pore size include a charge flux matching condition that sets the potential change at the pore-SDL transition region;
\item The proposed model predicts the potential and charge density profiles inside a nanopore. The predicted profiles using Eqs. (\ref{Eq: psival}), (\ref{Eq: rhoval}) and (\ref{Eq: final_sol_full}) show quantitative agreement with the results from DNS even for moderate applied potentials;
\item {Physical insight into the mechanisms setting capacitance and charging timescale of pores with arbitrary sizes is obtained: the influence of electromigration and charge diffusion is quantified;}
\item Electrical-double-layer charging for arbitrary pore sizes can be represented in the form of a transmission line circuit, but with the inclusion of a time-dependent interfacial capacitance. To mitigate this complexity, Eq. (\ref{Eq: chargeflux}) should be utilized to calculate charge flux; 
\item The electrode capacitance derived from an average of the total charge stored in the pore is able to capture { some} effects of pore size on pore capacitance reported in the literature \cite{simon2008materials,garcia2015constant}.
\end{itemize}

{Our methodology provides valuable insight into the effects of electromigration and diffusion in double-layer charging for arbitrary pore sizes. For thin double layers, electromigrative flux controls the charge flux and the charging timescale. In contrast, for overlapping double layers, the diffusive and electromigrative fluxes cooperate, enhancing the total charge flux. Yet, this increase in flux for overlapping double layers is less than the corresponding boost in charge density. This leads to a longer charging timescale in narrow pores, and thus a trade-off between charge stored and charging timescale. We also report a simplified model example of the influence of double layer thickness over electrode charging for non-interacting pores via phenomenological averages under a proposed log-normal distribution. In this case study, the same single-pore proportional increases of capacitance and charging timescale manifest in a distribution of isolated pores, predicting a constant electrode power density regardless of  relative pore size.}

\par{}Our approach can be extended to studying hybrid supercapacitors by adding reactions to the boundary conditions. The perturbation expansion analysis proposed here can also be utilized for asymmetric ionic valences \cite{gupta2018electrical} and diffusivities \cite{kim2019characterization,khair2020breaking,amrei2020perturbation,balu2021thin}, scenarios which are commonly observed in electrochemical devices. Finally, to compare directly with cyclic voltammetry data obtained from experiments, a similar approach can also be employed for higher and/or time-dependent applied potentials.

\section*{Conflicts of Interest}
There are no conflicts to declare.

\section*{Acknowledgements}
The authors would like to acknowledge the helpful input provided by Howard A. Stone, Daniel K. Schwartz, Adam Holewinski, Gesse Roure, Nathan Jarvey, and {anonymous Referee \#2}. P.J.Z. would like to acknowledge the support of a project that has received funding from the European Union’s Horizon 2020 research and innovation program under the Marie Skłodowska-Curie grant agreement No. 847413 and was a part of an international co-financed project founded from the program of the Minister of Science and Higher Education entitled “PMW” in the years 2020–2024; agreement No. 5005/H2020-MSCA-COFUND/2019/2. The authors acknowledge that the simulations reported in this contribution were performed using the
Princeton Research Computing resources at Princeton University which is consortium of groups including the Princeton Institute for Computational Science and Engineering and the Princeton University Office of Information Technology’s Research Computing department.

\bibliography{apssamp}

\appendix

\setcounter{figure}{0}
\renewcommand\thefigure{A\arabic{figure}}

\section{Transmission Line Circuit Resistances}

In order to complete the transmission line circuit description in Fig. 9a, we derive the expressions of dimensional SDL and transition-region resistances. We start with the SDL resistance, which is given by Ohm's law as
\begin{equation}
    \tilde{R}_s=\dfrac{2\ell_p\lambda^2}{\varepsilon D_pA_p}\dfrac{\Psi_\mathrm{left}}{J_\mathrm{left}}.
\end{equation}
Utilizing Eqs. (\ref{Eq: dim_Ns}) and (\ref{Eq: Psilin}), we have
\begin{equation}
    \tilde{R}_s=\dfrac{\ell_s\lambda^2}{\varepsilon D_sA_s}.
    \label{eq:Rsdim}
\end{equation}

\noindent From Eq. (\ref{Eq: Psi1Psi2}), the potential difference across the transition region is given by
\begin{equation}
    \Psi_\mathrm{left}-\Psi_\mathrm{right}=\dfrac{\Psi_\mathrm{right}-\Psi_D}{I_0(\frac{a_p}{\lambda})-1}.
\end{equation}
Using Eq. (\ref{Eq: final_sol}), it can be written as
\begin{equation}
    \resizebox{\linewidth}{!}{$\Psi_\mathrm{left}-\Psi_\mathrm{right}=-\dfrac{\Psi_D}{I_0(\frac{a_p}{\lambda})}\left[1-2\sum_{n=1}^\infty\dfrac{\sin 2\kappa_n}{2\kappa_n+\sin 2\kappa_n}\exp(-\kappa_n^2T)\right].$}
    \label{eq:psilpsir}
\end{equation}
Therefore, the dimensional transition-region resistance follows from Eq. (\ref{Eq: chargeflux}) as
\begin{equation}
    \tilde{R}_\mathrm{t}=\dfrac{\ell_p\lambda^2}{\varepsilon D_pA_p}\frac{1-2\sum_{n=1}^\infty\frac{\sin 2\kappa_n}{2\kappa_n+\sin 2\kappa_n}\exp(-\kappa_n^2T)}{2\mathrm{Bi}\,I_0(\frac{a_p}{\lambda})\sum_{n=1}^\infty\frac{\sin 2\kappa_n}{2\kappa_n+\sin 2\kappa_n}\exp(-\kappa_n^2T)}.
    \label{eq:Rtdim}
\end{equation}
We choose the scale $\frac{\ell_p\lambda^2}{\varepsilon D_pA_p}$ for the resistances, such that their dimensionless expressions read
\begin{equation}
    R_s=\dfrac{1}{\mathrm{Bi}}
\end{equation}
and
\begin{equation}
    R_t=\frac{1-2\sum_{n=1}^\infty\frac{\sin 2\kappa_n}{2\kappa_n+\sin 2\kappa_n}\exp(-\kappa_n^2T)}{2\mathrm{Bi}\,I_0(\frac{a_p}{\lambda})\sum_{n=1}^\infty\frac{\sin 2\kappa_n}{2\kappa_n+\sin 2\kappa_n}\exp(-\kappa_n^2T)}.
\end{equation}

\begin{figure}[h!]
    \centering
    \includegraphics[scale=0.4]{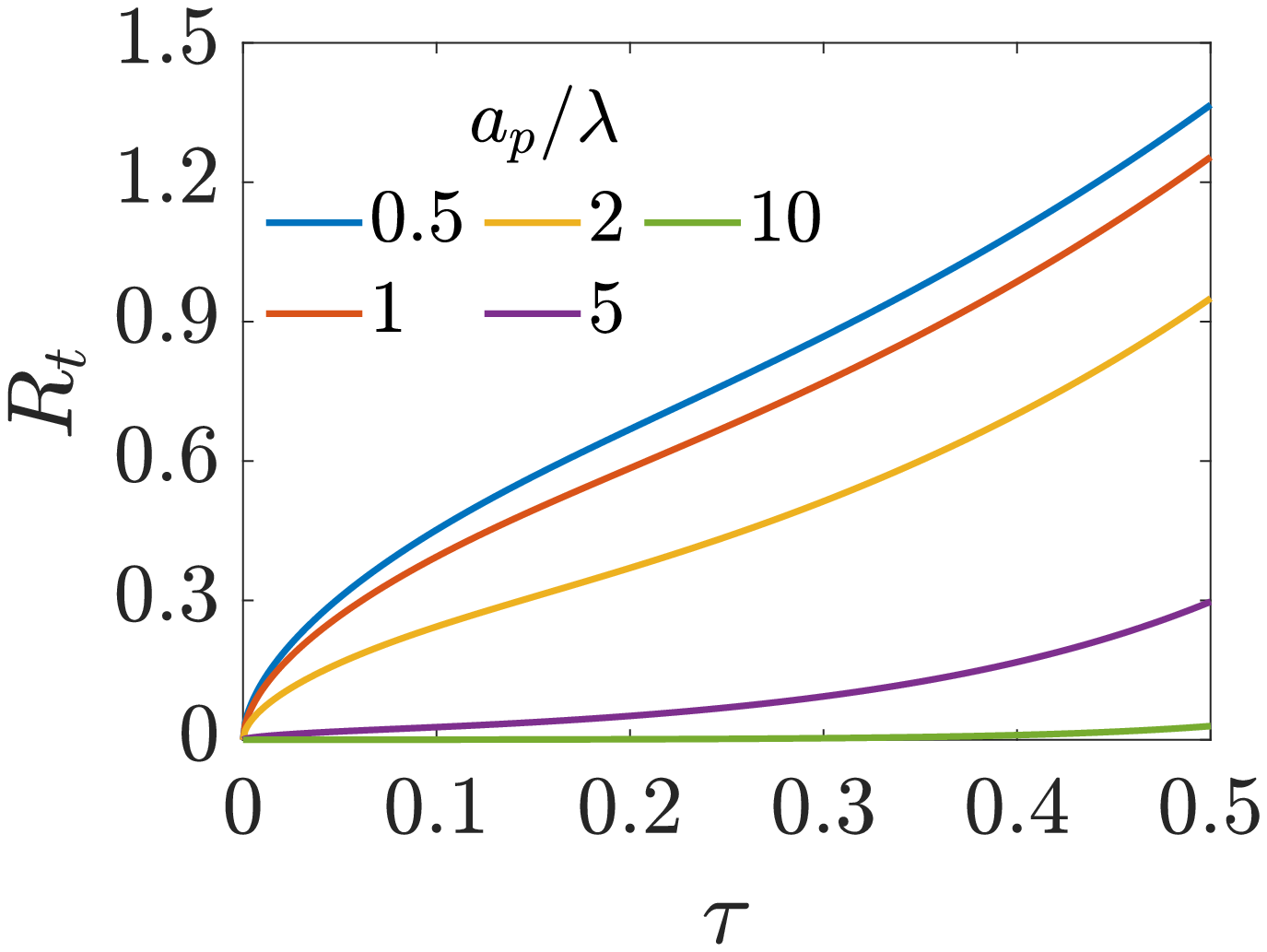}
    \caption{Resistance of the transition region as a function of time for different relative pore sizes. The resistance is larger for narrower pores and increases in time to maintain a potential difference (see Eq. (\ref{Eq: Psi1Psi2})) across the entrance region.}
    \label{fig:Rtr}
\end{figure}

Fig. (\ref{fig:Rtr}) shows a plot of the transition resistance over time for different relative pore sizes. As the relative pore size decreases, the resistance imposed by the transition region to the charge flux increases. This is due to the larger potential differences across the transition region for lower relative pore sizes, caused by their increased charge density.

\section{Areal Capacitance}

\begin{figure}
    \centering
    \includegraphics[scale=0.4]{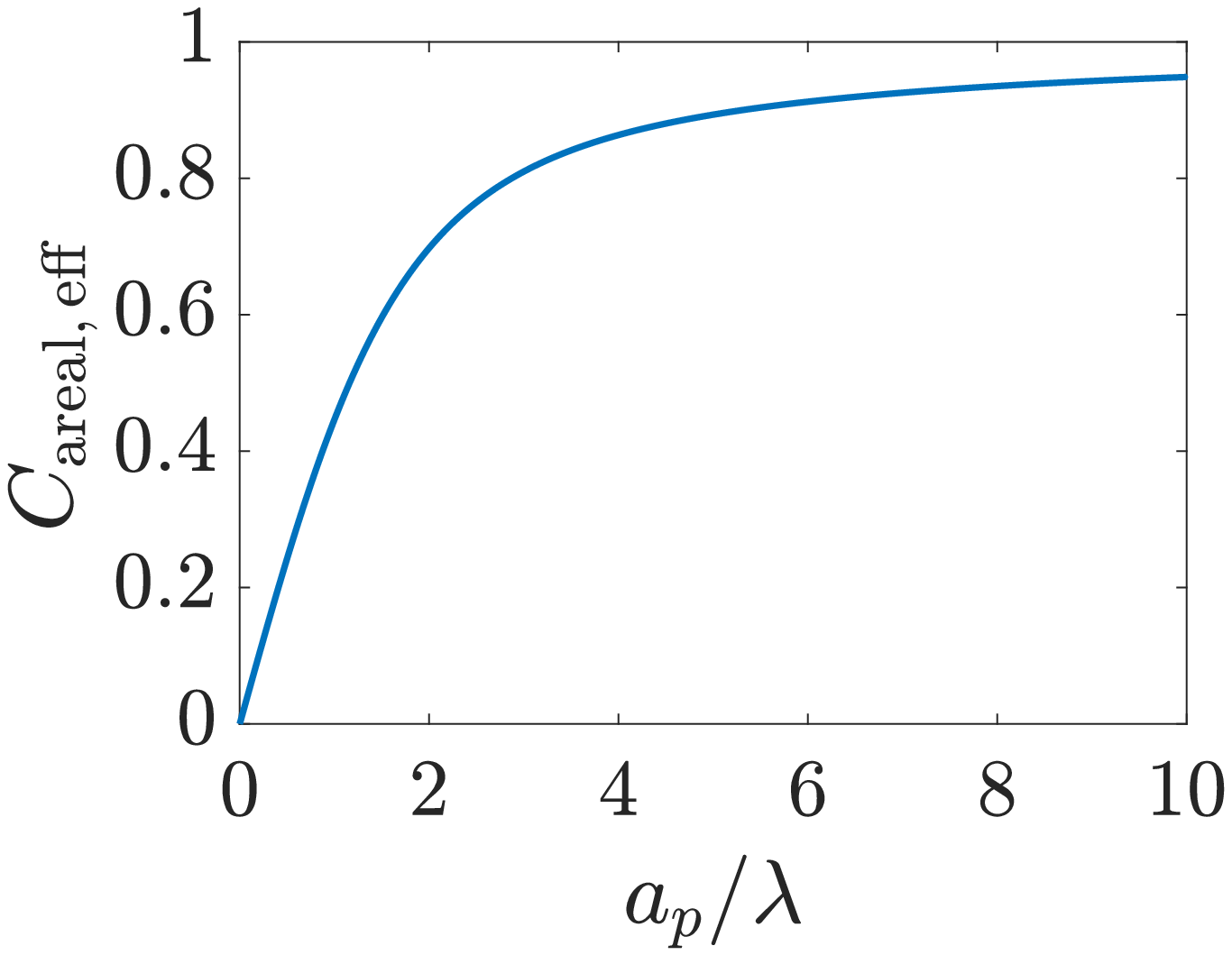}
    \caption{Dimensionless effective areal capacitance versus relative pore size. The increase in capacitance for larger pores predicted by the model developed in the current work accurately captures the trend observed in experiments for nanopores (Sec. III of Fig. 4 of Ref. \cite{simon2008materials}) and  qualitatively agrees with the results of Fig. 1 of Ref. \cite{garcia2015constant}.}
    \label{fig:capaeff}
\end{figure}

Most experiments report results on a per unit surface area basis, hence we briefly present our results for areal capacitance here in order to compare them qualitatively to experiments. Denoting dimensionless areal capacitance by $C_{\textrm{areal, eff}}$ and scaling it by $\frac{\varepsilon}{\lambda}$, it follows from Eq. (\ref{Eq: ceffvol}) that
\begin{equation}
    C_{\textrm{areal, eff}}=\frac{I_1(\frac{a_p}{\lambda})}{I_0(\frac{a_p}{\lambda})}.
\end{equation}
We plot this result in Fig. (\ref{fig:capaeff}).

\section{Electrode Average Properties} 

We note that the methodology presented here assumes  non-interacting pores and overlooks pore-network configuration, pore intersections, connectivity, among others. To extend single-pore analysis for a distribution of non-interacting pores, we first consider a continuous log-normal distribution of pore sizes,
\begin{equation}
    p\left(\frac{a_p}{\lambda}\right)=\dfrac{\lambda}{\sqrt{2\pi\sigma^2}a_p}\exp\left(-\dfrac{(\ln(\frac{a_p}{\lambda})-\mu)^2}{2\sigma^2}\right),
\end{equation}
where the parameters $\mu$ and $\sigma$ are given in terms of relative pore size average and variance by $\langle\frac{a_p}{\lambda}\rangle=\exp(\mu+\frac{\sigma^2}{2})$ and $\mathrm{Var}(\frac{a_p}{\lambda})=[\exp(\sigma^2)-1]\exp(2\mu+\sigma^2)$.

Next, we formally derive the averaging procedure employed in Fig. \ref{fig:ceffavg}. The average volumetric capacitance is defined as the total charge stored inside the electrode $Q$ divided by its total volume $V$ and applied potential $\Psi_D$, compatible with what experiments denote as volumetric capacitance \cite{boota2015graphene}
\begin{equation}
    C_{\mathrm{eff}}=\dfrac{Q}{\Psi_D V}.
    \label{eq:ceff}
\end{equation}
Charge is the integral of the charge density over the entire electrode volume,
\begin{equation}
    Q=\int_V\rho\,dV =\int_{V_e} \rho\,dV + \int_{V_p} \rho\,dV,
\end{equation}
where $V_e$ and $V_p$ are volumes of electrode material and pores, respectively. Furthermore, $\rho=0$ inside the electrode material (an ideal conductor). Therefore, the results integral can be written as a sum over all pores
\begin{equation}
    Q=\sum_{i=1}^N\int_{V_{p,i}}\rho\,dV=\sum_{i=1}^N q_i,
    \label{eq:qi}
\end{equation}
where $q_i$ is the charge of a pore with volume $V_{p,i}$, both known from the single-pore model. $N$ represents the total number of pores. We insert Eq. (\ref{eq:qi}) into Eq. (\ref{eq:ceff}) and take averages, denoted by $\langle\rangle$, over an ensemble of sample electrode pore configurations drawn from the same pore-size distribution. In this way, we have
\begin{equation}
    \langle C_{\mathrm{eff}}\rangle=\dfrac{ \langle \sum_{i=1}^N  q_i \rangle}{\psi_D V} = \dfrac{N  \langle q_i \rangle}{\psi_D V}.
    \label{eq:ceffav}
\end{equation}
For a non-interacting pore model, the average charge of the i-th pore  only depends on its relative pore size, i.e.,
\begin{equation}
    \langle q_i\rangle=\langle q\rangle=\int_0^\infty qp\left(a_p\right)\,d\left(a_p\right).
    \label{eq:qav}
\end{equation}
Substituting Eq. (\ref{eq:qav}) into Eq. (\ref{eq:ceffav}), we have
\begin{equation}
    \langle C_{\mathrm{eff}}\rangle=\dfrac{N}{V}\dfrac{\int_0^\infty qp\left(a_p\right)\,da_p}{\Psi_D}.
    \label{eq:ceffav2}
\end{equation}
Next, we multiply and divide the result by the average pore volume of the electrode
\begin{equation}
    \langle V_p\rangle=\int_0^\infty \pi a_p^2\ell_p p\left(a_p\right)\,da_p,
\end{equation}
which yields
\begin{equation}
    \langle C_{\mathrm{eff}}\rangle=\dfrac{N\langle V_p\rangle}{V}\dfrac{\int_0^\infty qp\left(a_p\right)\,da_p}{\Psi_D\int_0^\infty \pi a_p^2\ell_p p\left(a_p\right)\,da_p}.
    \label{eq:ceffav3}
\end{equation}
Lastly, we write the charge of a pore in terms of its volumetric effective capacitance, $q=C_\mathrm{eff}\Psi_D\pi a_p^2\ell_p$, change variables in the integrals from $a_p$ to $a_p/\lambda$ and define the electrode porosity $\phi=N\langle V_p\rangle/V$ to get
\begin{equation}
    \dfrac{\langle C_\mathrm{eff}\rangle}{\phi}=\dfrac{\int_0^\infty C_\mathrm{eff}\,p\left(\frac{a_p}{\lambda}\right)\left(\frac{a_p}{\lambda}\right)^2\,d\left(\frac{a_p}{\lambda}\right)}{\int_0^\infty p\left(\frac{a_p}{\lambda}\right)\left(\frac{a_p}{\lambda}\right)^2\,d\left(\frac{a_p}{\lambda}\right)}.
\end{equation}

Next, we propose a definition of timescale for a distribution of pore sizes which reduces to a single-pore solution for a monodisperse distribution. To this end, note that Eq. (\ref{Eq: dim_rho_ztau}) can be integrated over the volume of a pore to give, using Eq. (\ref{Eq: dim_Np}) and Eqs. (\ref{eq: bcsend}) for a blocking electrode, to give
\begin{equation}
    \dfrac{dq_i}{d\tau}=J_\mathrm{right,i}A_p,
    \label{eq:qdot}
\end{equation}
where $J_\mathrm{right}=J_Z|_{Z=0^+}$ and the index $i$ denotes the i-th pore. Integrating over time and performing the change of variables $T=\tau/\tau_{c,i}$, we have
\begin{equation}
    q_i(\tau)=\tau_{c,i}A_p\int_0^{T(\tau)} J_\mathrm{right,i}\,dT.
    \label{eq:qtau}
\end{equation}
Using (\ref{eq:qdot}) and  (\ref{eq:qtau}), we find that
\begin{equation}
    \tau_{c,i}=\dfrac{J_\mathrm{right,i}(\tau=0)}{\int_0^\infty J_\mathrm{right,i}(T(\tau))\,dT} \dfrac{q_i(\tau\to\infty)}{\frac{dq_i}{d\tau}|_{\tau=0}}.
\end{equation}
Our motivation in generalizing this is to construct a measure of the total charging time of the electrode, not of the individual pores. Therefore, noting that neither the initial value of the flux nor its integral on time depend on pore size -- see Eq. (\ref{Eq: chargeflux}) -- we opt to define the electrode charging timescale as
\begin{equation}
    \langle\tau_{c}\rangle=\dfrac{J_\mathrm{right}(\tau=0)}{\int_0^\infty J_\mathrm{right}(T(\tau))\,dT}\dfrac{\langle q(\tau\to\infty)\rangle}{\langle\frac{dq}{d\tau}|_{\tau=0}\rangle},
\end{equation}
averaged over the ensemble of pore configurations, which explicitly takes the form
\begin{equation}
    \langle\tau_c\rangle=\dfrac{J_\mathrm{right}(\tau=0)}{\int_0^\infty J_\mathrm{right}(T(\tau))\,dT}\dfrac{\int_0^\infty q(\tau\to\infty)p\left(a_p\right)\,d\left(a_p\right)}{\int_0^\infty \frac{dq}{d\tau}|_{\tau=0}\,p\left(a_p\right)\,d\left(a_p\right)}.
\end{equation}
Using Eqs. (\ref{eq:qdot}) and  (\ref{eq:qtau}) where $A_p=\pi a_p^2$ and changing variables from $a_p$ to $a_p/\lambda$, we have
\begin{equation}
\langle\tau_c\rangle=\dfrac{\int_0^\infty \tau_c\,p\left(\frac{a_p}{\lambda}\right)\left(\frac{a_p}{\lambda}\right)^2\,d\left(\frac{a_p}{\lambda}\right)}{\int_0^\infty p\left(\frac{a_p}{\lambda}\right)\left(\frac{a_p}{\lambda}\right)^2\,d\left(\frac{a_p}{\lambda}\right)}.
\end{equation}
Through this approach we define a physically relevant average of charging timescale that is consistent with the single-pore definition.

\end{document}